\documentclass{article}

\usepackage{arxiv}

\usepackage{authblk}
\usepackage{hyperref} 
\usepackage{csquotes} 
\usepackage{url}      
\usepackage{graphicx}
\usepackage[numbers,sort]{natbib}
\usepackage{subfiles}
\usepackage{tabularx}
\usepackage{etoolbox}
\usepackage{multirow}
\usepackage{enumitem}
\usepackage{amssymb}
\usepackage{amsmath}
\usepackage{amsthm}
\usepackage{xspace}
\usepackage{float}
\usepackage{wrapfig}
\usepackage{booktabs}  
\usepackage{amsfonts}  
\usepackage{nicefrac}  
\usepackage{microtype} 
\usepackage{color, soul}
\usepackage{pifont}
\usepackage{nth}

\usepackage[framemethod=TikZ]{mdframed}

\usepackage{amsmath,amssymb,amsthm,amsfonts,latexsym,bbm,xspace,graphicx,float,mathtools,mathdots, multirow, multicol}

\ifdefined\ShowNotes
  \newcommand{\colornote}[3]{{\color{#1}\bf{#2 #3}\normalfont}}
\else
  \newcommand{\colornote}[3]{}
\fi

\definecolor{darkred}{rgb}{0.7,0.1,0.1}
\definecolor{darkgreen}{rgb}{0.1,0.5,0.1}
\definecolor{cyan}{rgb}{0.7,0.0,0.7}
\definecolor{dblue}{rgb}{0.2,0.2,0.8}
\definecolor{maroon}{rgb}{0.76,.13,.28}
\definecolor{burntorange}{rgb}{0.81,.33,0}
\definecolor{royalpurple}{rgb}{0.47,.31,0.66}

\ifdefined\ShowNotes
  
\else
  
\fi

\definecolor{dkgreen}{rgb}{0,0.6,0}
\definecolor{gray}{rgb}{0.5,0.5,0.5}
\definecolor{light-gray}{gray}{0.95}
\definecolor{mauve}{rgb}{0.58,0,0.82}
\definecolor{backcolour}{rgb}{0.95,0.95,0.92}

\usepackage{graphicx}
\usepackage{subfigure}
\usepackage{wrapfig,lipsum,booktabs}
\usepackage{threeparttable}
\usepackage{makecell}
\usepackage{hyperref}
\usepackage{url}
\usepackage{algorithm}
\usepackage{algpseudocode}
\usepackage{hyperref}
\usepackage{cleveref}
\usepackage{changepage}
\usepackage{booktabs}

\usepackage{epstopdf}
\usepackage{graphicx}
\usepackage{textcomp} 
\usepackage{scalerel} 

\usepackage{listings}
\usepackage{color}
\definecolor{dkgreen}{rgb}{0,0.6,0}
\definecolor{gray}{rgb}{0.5,0.5,0.5}
\definecolor{mauve}{rgb}{0.58,0,0.82}

\lstdefinelanguage{CUDACPP}{
  language=C++,
  morekeywords={__global__, __host__, __device__, __shared__, blockIdx, blockDim, threadIdx, gridDim},
  morecomment=[l][\color{magenta}]{\#},
}

\lstdefinestyle{pythonstyle}{
  language=Python,
  frame=tb,
  aboveskip=3mm,
  belowskip=3mm,
  showstringspaces=false,
  columns=flexible,
  basicstyle={\ttfamily\footnotesize},
  numbers=none,
  numberstyle=\footnotesize\color{gray},
  keywordstyle=\color[rgb]{0.13,0.29,0.53},
  commentstyle=\color[rgb]{0.13,0.55,0.13},
  stringstyle=\color[rgb]{0.31,0.60,0.02},
  breaklines=true,
  breakatwhitespace=true,
  tabsize=4
}

\usepackage{amsmath,amsfonts,bm}

\def\eqref#1{equation~\ref{#1}}

\def\1{\bm{1}}

\DeclareMathAlphabet{\mathsfit}{\encodingdefault}{\sfdefault}{m}{sl}
\SetMathAlphabet{\mathsfit}{bold}{\encodingdefault}{\sfdefault}{bx}{n}

\usepackage{listings}
\definecolor{codegreen}{rgb}{0,0.6,0}
\definecolor{codegray}{rgb}{0.5,0.5,0.5}
\definecolor{codepurple}{rgb}{0.58,0,0.82}
\definecolor{backcolour}{rgb}{1,1,1}

\usepackage{listings}
\usepackage{algorithm}
\algrenewcommand\algorithmicrequire{\textbf{Input:}}
\algrenewcommand\algorithmicensure{\textbf{Output:}}

\newcommand{\ShowNotes}{}

\newlength{\defbaselineskip}
\newcolumntype{Y}{>{\hsize=.7\hsize}X}
\newcolumntype{Z}{>{\hsize=1.3\hsize}X}

\makeatletter
\def\@copyrightspace{\relax}
\makeatother

\makeatletter
\def\@myauthornotes{}
\def\myauthornote#1{%
  \if@ACM@anonymous\else
    \g@addto@macro\addresses{}%
    \g@addto@macro\@myauthornotes{%
      \stepcounter{footnote}\footnotetext{#1}}%
  \fi}
\makeatother

\newcommand{\name}{\textsc{ParallelKittens}}
\newcommand{\shortname}{\textsc{PK}}
\newcommand{\tk}{\textsc{ThunderKittens}}
\newcommand{\shorttk}{\textsc{TK}}

\title{\name: Systematic and Practical Simplification\\of Multi-GPU AI Kernels}

\author[]{Stuart H. Sul}
\author[]{Simran Arora}
\author[]{Benjamin F. Spector}
\author[]{Christopher Ré}
\affil[]{Department of Computer Science, Stanford University}
\affil[]{\{\texttt{ssul,simarora,bfs,chrismre\}@stanford.edu}}

\begin{document}

\maketitle

\begin{abstract}

Inter-GPU communication has become a major bottleneck for modern AI workloads as models scale and improvements in hardware compute throughput outpace improvements in interconnect bandwidth.
Existing systems mitigate this through compute-communication overlap but often fail to meet theoretical peak performance across heterogeneous workloads and new accelerators.
Instead of operator-specific techniques, we ask whether a small set of simple, reusable principles can systematically guide the design of optimal multi-GPU kernels.
We present {\name} ({\shortname}), a minimal CUDA framework that drastically simplifies the development of overlapped multi-GPU kernels. {\shortname} extends the {\tk} framework and embodies the principles of multi-GPU kernel design through eight core primitives and a unified programming template, derived from a comprehensive analysis of the factors that govern multi-GPU performance---data-transfer mechanisms, resource scheduling, and design overheads. We validate {\shortname} on both Hopper and Blackwell architectures. With fewer than 50 lines of device code, {\shortname} achieves up to $2.33\times$ speedup for data- and tensor-parallel workloads, $4.08\times$ for sequence-parallel workloads, and $1.22\times$ for expert-parallel workloads.

\end{abstract}

\section{Introduction}
\label{sec:introduction}

\begin{figure}[t]
    \centering
    \includegraphics[
        width=\textwidth
    ]{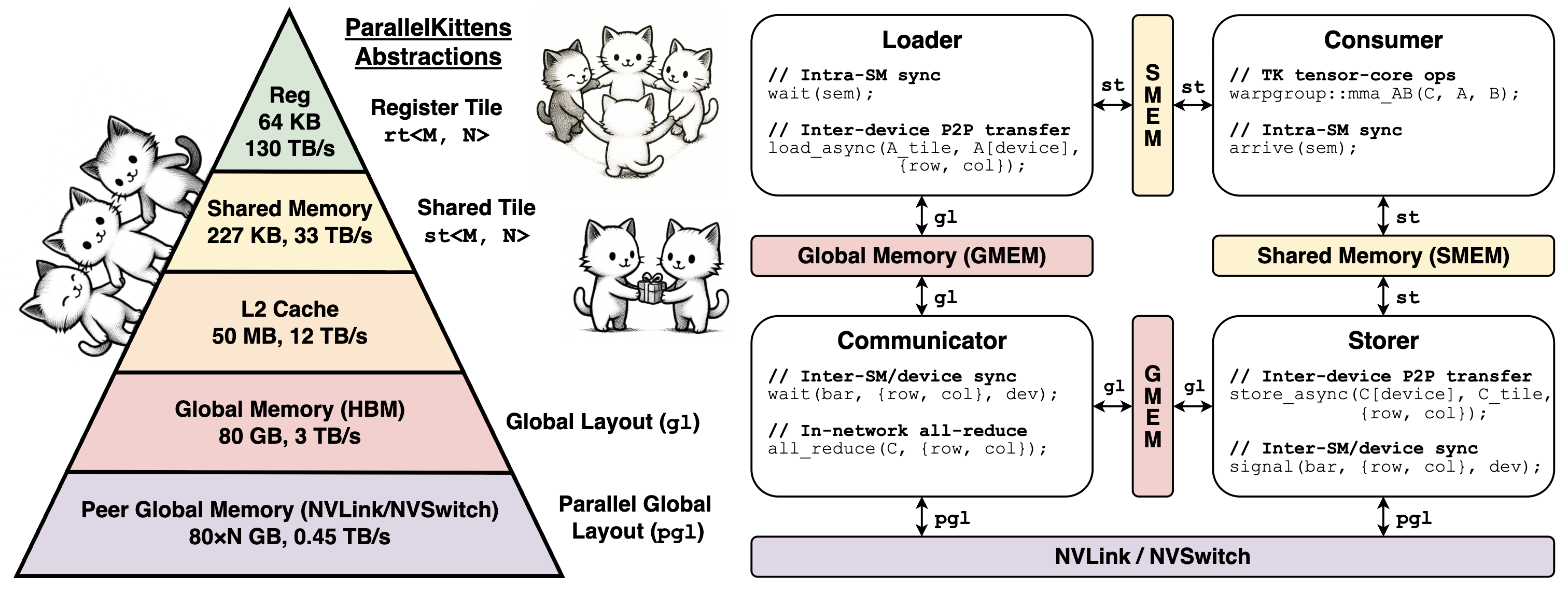}
    \caption{We study the principles for high performance multi-GPU kernels and introduce {\name} (\shortname), an opinionated collection of programming primitives to encapsulate these principles. 
    The GPU memory hierarchy and corresponding {\shortname} abstractions are shown on the left (Section~\ref{sec:pk-data-structure}), and the {\shortname} program template with its key multi-GPU kernel components is shown on the right (Section~\ref{sec:pk-program-template}).}
    \label{fig:pull}
\end{figure}

A few years ago, GPU compute utilization was often limited by \textit{intra-GPU} memory access. However, IO-aware algorithms like FlashAttention \cite{22-FA1}, domain-specific languages (DSLs) that support efficient mapping of operators to hardware \cite{19-triton, 24-cute-dsl, 25-thunderkittens}, and the continued scaling of AI models have have left \textit{inter-GPU} communication as the primary remaining bottleneck. Even with high-speed interconnects like NVLink \cite{nvlink-nvswitch} and compute-friendly phases like prefill, communication can occupy over $50\%$ of execution time in large language model (LLM) workloads, leaving GPU compute idle \cite{24-flux}. The problem is compounded by the relatively slow improvements in communication hardware: from the Nvidia A100 \cite{20-ampere} to the B200 \cite{25-blackwell}, BF16 tensor core performance improved by $7.2 \times$ and High Bandwidth Memory (HBM) bandwidth by $5.1 \times$, while intra-node communication (NVLink) improved by only $3 \times$ and inter-node (PCIe/InfiniBand) by just $2 \times$.

To mitigate communication overhead, prior methods \textit{overlap} inter-GPU communication with intra-GPU computation for common operators like General Matrix Multiplication (GEMM), attention, and Mixture-of-Experts (MoE) layers \cite{24-ring-attention, 24-flux, 25-deepep, 25-triton-dist, 25-comet, 25-flashdmoe}. These approaches reduce non-overlapped communication time in data, tensor, sequence, and expert parallelism \cite{19-megatron, 25-torchtitan}, which are common strategies for distributing industry-scale training and inference across many GPUs. However, prior works either (i) rely on bespoke kernels for specific AI operators and depend on complex low-level primitives (e.g., CUTLASS, NVSHMEM, Linux IPC), (ii) employ compiler-based approaches that fail to adapt to new accelerators---occasionally generating kernels slower than non-overlapped baselines---or (iii) utilize off-the-shelf libraries, resulting in up to $4.08\times$ slower performance than hand-tuned implementations.

As hardware shifts toward unified multi-GPU systems---illustrated by Nvidia’s roadmap from NVL72 to NVL144 (2026) and NVL576 (2027) \cite{25-nvl-roadmap}---we would need simple, general principles and programming primitives that enable peak-performance multi-GPU operations. In this work, we identify three key principles for designing efficient multi-GPU kernels and analyze each in detail (Section~\ref{sec:analysis}).

\begin{enumerate}[itemsep=0.1pt,topsep=0pt,leftmargin=*]
    \item \textbf{Transfer mechanism.} Inter-GPU networking relies on three mechanisms---copy engines, tensor memory accelerators (TMA), and register-level instructions---that differ in maximum bandwidth, effective message granularity, supported functionality, and compute occupancy. Understanding these trade-offs and choosing the right mechanism is crucial for peak performance. For instance, copy engines achieve the highest efficiency (81\% of theoretical maximum) but require large messages ($\geq$ 256 MB) for saturation. TMA attains near-peak throughput (74\%) with only 2 KB messages (Figure~\ref{fig:1gb-transfer-msg-size}). Register-level instructions operate efficiently at a 128 B granularity but need about 76 streaming multiprocessors (SMs) to saturate bandwidth (70\%), whereas TMA needs 15 (Figure~\ref{fig:nvl-saturation}). However, only register-level instructions support in-network reduction. Existing systems do not capture these trade-offs; for instance, Triton Distributed, Flux, and CUTLASS rely on the copy engine for intra-node all-gather GEMM, becoming slower than the non-overlapped baseline on smaller matrix sizes (Figure~\ref{fig:ag-gemm-h100}).

    \item \textbf{Scheduling.} The distribution of compute and communication work across SMs must be chosen based on workload characteristics. We identify \textit{inter-SM} and \textit{intra-SM} overlapping as the two primary scheduling strategies, trading off compute utilization and communication versatility. Intra-SM overlapping is preferred when computation and communication granularities align; for example, in GEMM reduce-scatter, intra-SM overlapping outperforms inter-SM by $1.2\times$. In contrast, inter-SM overlapping enables communication patterns that can significantly reduce transfer size. For instance, leveraging in-network reduction through inter-SM overlapping achieves a $3.62\times$ performance improvement for GEMM all-reduce (Figure~\ref{fig:inter-sm-schedules}) and $1.57\times$ for all-gather GEMM. No prior work explores both scheduling strategies; existing methods either rely on a single type or omit device-side overlapping altogether, thereby failing to generalize (e.g., applying the Flux intra-SM overlapping design to GEMM all-reduce would lead to the slowdown above).    

    \item \textbf{Design overheads.} Widely used communication libraries (e.g., NCCL, NVSHMEM) encapsulate design choices---specifically in synchronization and buffering---that favor simplicity over performance. We show that the choices in prior libraries can cause over $1.7\times$ performance loss in pure communication kernels (e.g., all-reduce) and up to $4.5\times$ higher communication latency. By adopting a design that enables explicit user control over memory allocation and synchronization, these overheads can be substantially reduced.
\end{enumerate}

Building on these insights, we introduce {\textbf{\name}} ({\shortname}), an opinionated collection of C++ embedded programming primitives that extends the {\tk} ({\shorttk}) framework \cite{25-thunderkittens} (Section~\ref{sec:abstractions}). {\shortname} exposes only the most efficient transfer mechanisms for each functionality (e.g., TMA for point-wise communication, register operations for in-network acceleration), provides minimal synchronization primitives and a general program template that simplifies achieving both inter- and intra-SM overlapping scheduling, and offers full control over performance-critical components (e.g., NVLink transfers) while abstracting away non-essential multi-GPU complexities (e.g., inter-process communication and virtual memory exchange).

We validate {\shortname} across diverse parallel AI workloads on both Hopper and Blackwell architectures, including data, tensor, sequence, and expert parallelism (i.e., fused parallel GEMMs, distributed attention variants, and MoEs). Compared with the strongest baselines, {\shortname} achieves up to $2.33\times$ higher compute throughput (FLOP/s) for data- and tensor-parallel workloads, $4.08\times$ for sequence-parallel workloads, and $1.22\times$ for expert-parallel workloads, effectively reducing non-overlapped communication time down to $1\%$, $9\%$, and $15\%$, respectively. {\shortname} matches the performance of the strongest hand-optimized kernels (Flux, Comet, CUTLASS), outperforms compiler-based approaches (Triton Distributed) by $1.07$--$5.63\times$, and surpasses communication library-based approaches (xDiT, YunChang) by $1.01$--$4.08\times$ across varying problem sizes.

Each {\shortname} kernel required fewer than 50 lines of additional device code beyond the original single-GPU GEMM or attention kernels. The complete implementation of {\shortname}, including its kernels, is fully open-sourced and is currently being adopted at Cursor for large-scale in-house training.

To summarize, our contributions are:

\begin{itemize}[itemsep=0.1pt,topsep=0pt,leftmargin=*]
    \item A detailed analysis of multi-GPU programming that decomposes performance into interpretable factors (transfer mechanisms, scheduling strategies, and design overheads) and validates each with microbenchmarks.
    \item {\name}, a minimal collection of multi-GPU primitives and a unified programming template that extends the familiar {\tk} framework.
    \item Kernels built with {\name} that match or surpass hand-optimized kernel performance while substantially reducing code complexity.
\end{itemize}

\section{Background}
\label{sec:background}

In this section, we provide background on modern datacenter-grade GPUs and review prior efforts on optimizing multi-GPU AI kernels.\footnote{Unless otherwise specified, we use the Nvidia HGX H100 \cite{22-hgx-h100} platform with 8$\times$H100 80GB SXM GPUs, 4\textsuperscript{th} generation NVLink/NVSwitch, and 5\textsuperscript{th} generation PCIe as our running example; however, the principles extend to other modern platforms (e.g., Blackwell architecture) and hardware vendors (e.g., AMD).}

\subsection{GPU Architecture}

A GPU kernel loads data from HBM, performs computation, and writes the results back to HBM. Multi-GPU kernels distribute the workload across multiple GPUs and access the HBMs of all devices.

\paragraph{GPU hierarchy.} GPU kernels execute tens of thousands of hardware \textit{threads} in parallel across over a hundred \textit{streaming multiprocessors} (SMs). Memory farther from the SM provides greater capacity at higher latency. Each SM contains 64~KB of registers private to individual threads and accessible every clock cycle. Threads are organized into \textit{thread blocks}, each executing on a single assigned SM. Threads in a thread block communicate via 227~KB of shared memory (SMEM), a per-SM on-chip SRAM offering up to 33~TB/s of bandwidth. All threads share a 50~MB L2 cache ($\approx$12~TB/s) connected to 80~GB HBM (3~TB/s). Threads can also access \textit{peer GPU HBM} over NVLink (450~GB/s unidirectional), enabling multi-GPU kernel development.

\paragraph{GPU networking.} Multi-GPU systems rely on a hierarchy of interconnects. \textit{PCIe} (64~GB/s) is the channel for CPU-to-GPU (e.g., kernel launches, host-initiated transfers) and multinode communication over InfiniBand/TCP. \textit{NVLink} (450~GB/s) provides point-to-point connections between GPUs and the NVSwitch; \textit{NVSwitch} interconnects all NVLink endpoints into a non-blocking fabric for full GPU-to-GPU communication. NVSwitch also supports in-network, off-device acceleration for multicast and reduction. Unless otherwise noted, all inter-GPU communication in this paper occurs via NVLink/NVSwitch.

\paragraph{Execution overlap.} GPUs contain various execution units specialized for different compute, memory, and communication operations. For compute, \textit{Tensor Cores} perform tiled matrix multiplications, while \textit{CUDA Cores} handle element-wise arithmetic. For memory, the \textit{Tensor Memory Accelerator} (TMA) performs bulk data transfers between SMEM and HBM and can be invoked asynchronously by a single thread. Alternatively, a per-GPU \textit{copy engine} (dedicated DMA unit) moves large contiguous regions of device memory independently of the SMs and is invoked from the host. 

Within an SM, threads can concurrently issue instructions to different execution units. Achieving optimal performance therefore depends on effectively overlapping their use to hide non-critical operations and maximize the throughput of critical ones. We distinguish \textit{inter-SM overlapping}, where entire SMs are dedicated almost exclusively to compute, memory, or communication tasks, from \textit{intra-SM overlapping}, where different warps or threads within the same SM concurrently drive compute, memory, or inter-GPU traffic. These resources saturate at different rates, creating opportunities for various overlapping strategies.

\subsection{Related Works}

We are inspired by the extensive amount of work that accelerates multi-GPU AI workloads.

\paragraph{Operator-specific kernels.} Many prior works hand-tune particular AI operators by overlapping computation and communication, e.g., TP-Async~\cite{24-tp-async}, Flux~\cite{24-flux}, Ring Attention~\cite{24-ring-attention}, DeepEP~\cite{24-dsv3}, Comet~\cite{25-comet}, FlashDMoE~\cite{25-flashdmoe}, and several distributed GEMM kernels from CUTLASS~\cite{cutlass}. These approaches employ techniques ranging from overlapping host-triggered copies with device kernels, to highly optimized on-device schedulers and device-initiated communication. While these systems deliver strong performance for specific targets, they demand complex implementations and offer limited reusable abstractions. For instance, FlashDMoE is optimized only for TF32 precision, with BF16/FP16 support still under development five months after its release. In contrast, {\shortname} distills general principles applicable across diverse workloads, achieving speedups comparable to hand-optimized kernels while simplifying implementation.

\paragraph{Scheduling frameworks.} Frameworks such as Megatron-LM~\cite{19-megatron}, FlexFlow~\cite{19-flexflow}, and NanoFlow~\cite{25-nanoflow} automate parallelization and scheduling, and are complementary to \shortname. These systems primarily orchestrate bulk collective operations (e.g., NCCL), which require synchronization before and after data transfers, and employ stream-level overlap. NanoFlow offers finer-grained scheduling by partitioning SMs among compute, memory, and network operations to saturate available bandwidth without full occupancy (i.e., inter-SM overlapping). However, achieving peak kernel performance also requires intra-SM warp specialization with device-initiated, tile-level transfers; {\shortname} provides that layer.

\paragraph{Multi-GPU programming primitives.} DSLs and libraries have been proposed to simplify multi-GPU kernel development~\cite{nccl, nvshmem, 25-triton-dist, 25-iris, 25-pallas}. Triton Distributed~\cite{25-triton-dist} and TileLink~\cite{25-tilelink} extend Triton~\cite{19-triton} with OpenSHMEM-style one-sided operations, enabling compiler-based generation of multi-GPU kernels. However, these approaches lack explicit workload distribution control (e.g., warp or SM specialization) needed for optimal overlap. Also, our benchmarks show that Triton Distributed, originally tuned for H800 GPUs, fails to adapt efficiently to other architectures such as H100s (Section~\ref{sec:experiments}). In contrast, {\shortname} provides a lightweight C++ layer that enables direct control over communication workload distribution, enabling arbitrary scheduling and optimization across Hopper and Blackwell GPUs. NCCLX~\cite{25-ncclx} complements {\shortname} by accelerating \textit{inter-node} collectives for large clusters ($\ge100$k GPUs), but does not exploit device-initiated asynchronous overlapping (via TMA) or in-network acceleration, both critical for fine-grained overlap with peak bandwidth utilization. 

\section{ParallelKittens}
\label{sec:parallel-kittens}

We present our analysis of the design tradeoffs of multi-GPU kernels and present \name.

\subsection{Analysis}
\label{sec:analysis}

We start with a general, high-level cost model that provides a roadmap for the analysis.

\subsubsection{Cost Model}
\label{sec:cost-model}

The objective of designing a multi-GPU kernel is to minimize its total wall-clock time \( T_{\text{kernel}} \), which reflects the combined cost of compute, memory, and communication operations. The key contributors include:

\begin{equation*}
\begin{split}
T_{\text{kernel}} = T_{\text{launch}}
    + \max(T_{\text{comp}},\, T_{\text{mem}},\, T_{\text{comm}})
    +\, T_{\text{non-overlap}} + T_{\text{sync}}
\end{split}
\label{eq:kernel_cost}
\end{equation*}

In this simple model, \( T_{\text{launch}} \) denotes the per-kernel launch cost, including host-side latency and per-thread block setup and teardown (e.g., tensor memory allocation and pipeline fill/drain phases). \( T_{\text{comp}} \), \( T_{\text{mem}} \), and \( T_{\text{comm}} \) represent the full-pipeline time spent on computation, memory access, and communication, respectively. Ideally, these components overlap so that the total time equals the maximum of the three, but \( T_{\text{non-overlap}} \) accounts for operations that cannot be overlapped. The cost of each component (e.g., \( T_{\text{comm}} \)) depends on the work size (\( S_{\text{comm}} \)) and achievable bandwidth (\( B_{\text{comm}} \)), i.e., \( T_{\text{comm}} = S_{\text{comm}} / B_{\text{comm}} \). Finally, \( T_{\text{sync}} \) captures the synchronization overhead across SMs or devices.

These costs are controlled by three design decisions: first, the specific \textbf{transfer mechanism} that we select to move data between GPUs (Section~\ref{sec:comm-mechanism}); second, the kernel \textbf{scheduling strategy} for overlapping computation and communication (Section~\ref{sec:scheduling}); and third, the communication abstraction’s \textbf{design choices}, including peer-memory allocation, management, and access (Section~\ref{sec:comm-overheads}).

\subsubsection{Transfer Mechanism}
\label{sec:comm-mechanism}

We now discuss the choice of communication mechanism.

\begin{wraptable}{r}{0.5\linewidth}
    \vspace{-2em}
    \caption{The observed NVLink bandwidth utilization (GB/s) when using all SMs to transfer 1GB of data, and its ratio to the theoretical maximum (450 GB/s for H100s, 900 GB/s for B200s).}
    \label{table:nvl-max-bw}
    \vspace{-0.5em}
    \begin{center}
    \begin{small}
    \begin{sc}
    \begin{tabular}{lcc}
        \toprule
        Method & H100 BW (Ratio) & B200 BW (Ratio) \\
        \midrule
        Copy Engine & 368.82 (82\%) & 726.13 (81\%) \\
        TMA Op & 350.01 (78\%) & 669.12 (74\%) \\
        Register Op & 342.68 (76\%) & 628.35 (70\%) \\
        \bottomrule
    \end{tabular}
    \end{sc}
    \end{small}
    \end{center}
    \vspace{-2em}
\end{wraptable}

\begin{figure}[t]
    \centering
    \includegraphics[
        width=0.5\textwidth
    ]{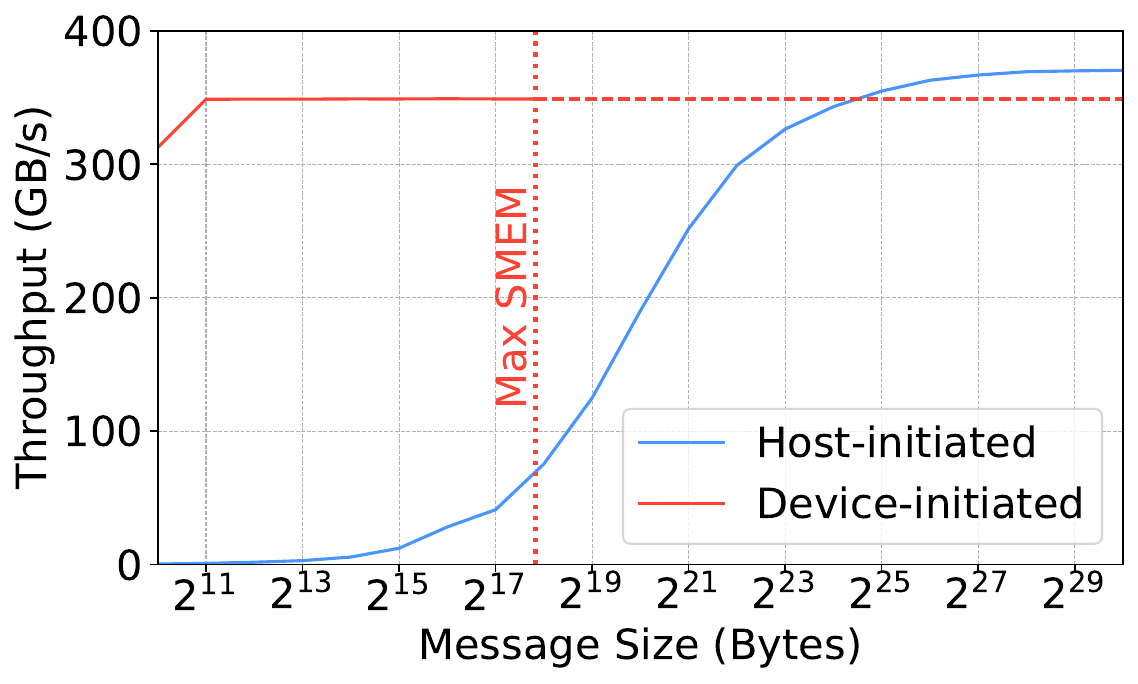}
    \caption{Observed memory bandwidth utilization for a 1 GB peer-to-peer transfer over NVLink. For device-initiated (TMA) transfers, the maximum supported message size is 227 KB; throughput values beyond this limit are held constant for visual comparison.}
    \label{fig:1gb-transfer-msg-size}
\end{figure}

\paragraph{Host versus device-initiated communication.} The per-GPU copy engine is host-initiated and supports only contiguous memory transfers. As shown in Table~\ref{table:nvl-max-bw}, it achieves the highest throughput for large, all-at-once data movements. However, when fine-grained communication is required (e.g., all-to-all communication in MoEs), performance degrades significantly because additional overhead is incurred for data rearrangement or repeated transfer invocations. Figure~\ref{fig:1gb-transfer-msg-size} illustrates this behavior. To sustain over 80\% bandwidth utilization, the transfer granularity must be at least 256~MB when using the copy engine, whereas device-side methods achieve comparable utilization with only 2~KB.

\begin{figure}[t]
    \centering
    \includegraphics[
        width=0.7\textwidth
    ]{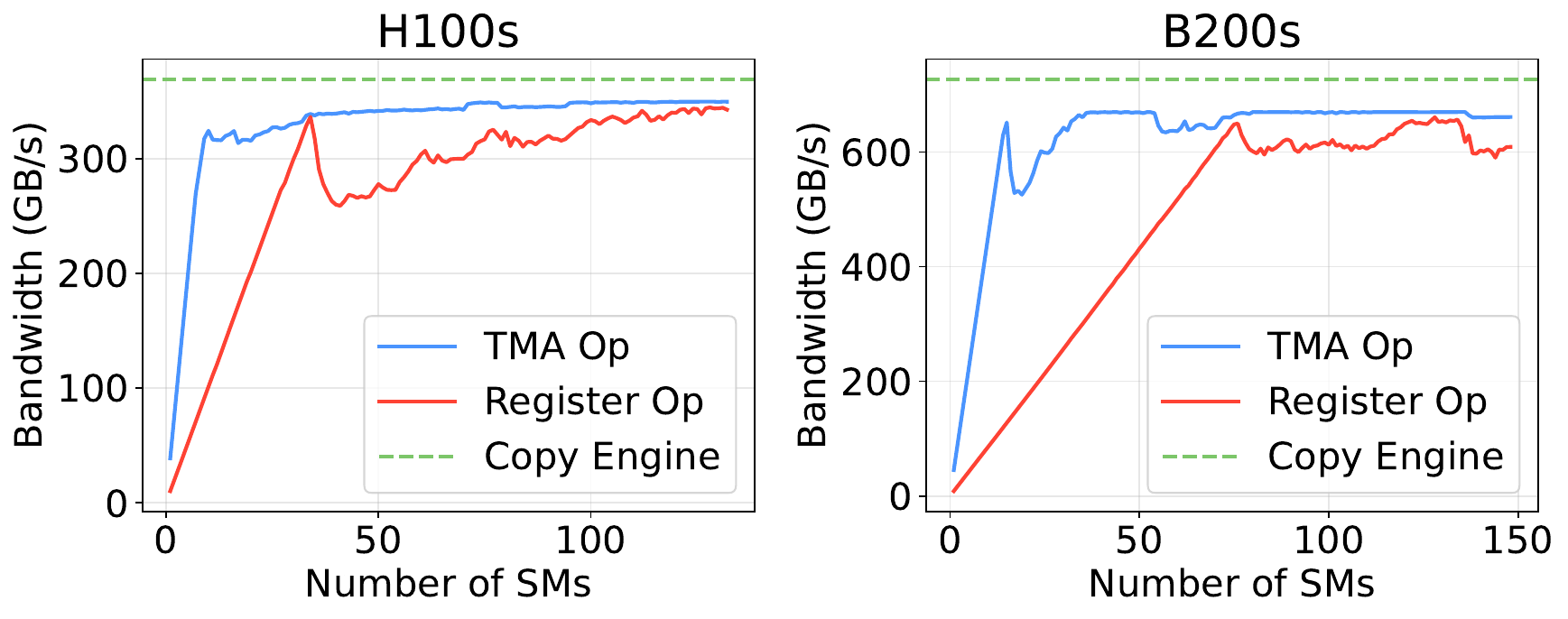}
    \caption{The number of SMs it takes to saturate NVLink Bandwidth, using different communication mechanisms.}
    \label{fig:nvl-saturation}
\end{figure}

Consequently, {\shortname} relies exclusively on device-side communication for the following reasons. First, host-initiated transfers are suitable primarily for large contiguous data blocks (e.g., weight movements in fully sharded data parallelism~\cite{zhao2023fsdp}). In such cases, overlapping computation and communication is often trivial: the host transfer and device kernel can be launched on separate streams without kernel modifications. Second, although the copy engine has the advantage of not occupying SM resources, only a small number of SMs are needed to saturate the interconnect bandwidth using device-initiated communication, as shown in Figure~\ref{fig:nvl-saturation}. Moreover, intra-SM overlapping enables computation to proceed concurrently with that communication. 

\paragraph{Device-initiated communication mechanisms.} There are two main mechanisms for device-initiated communication in modern datacenter-grade GPUs:

\begin{enumerate}[itemsep=0.1pt,topsep=0pt,leftmargin=*]
    \item The first is via the Tensor Memory Accelerator (TMA), which supports NVLink transfers and NVSwitch-accelerated broadcasts. A key advantage of TMA is that it can be launched \textit{asynchronously} by a \textit{single thread} without increasing register pressure, allowing other threads in the same SM to overlap the execution of compute or memory work (intra-SM overlap).
    \item The second is via plain register-level instructions (e.g., \texttt{ld}, \texttt{st}). As shown in Table~\ref{table:nvl-max-bw}, they are relatively inefficient, achieving about 70\% of the peak bandwidth on B200 GPUs. Because these instructions are synchronous and operate at the register level, saturating NVLink bandwidth requires full SM occupancy---thousands of threads issuing instructions concurrently---as well as higher register pressure and manual memory coalescing.
\end{enumerate}

\begin{wraptable}{r}{0.5\linewidth}
    \vspace{-2em}
    \caption{Different multi-GPU transfer mechanisms (copy engine, TMA, and register operations) and supported functionalities.}
    \label{table:comm-method-comparisons}
    \vspace{-0.5em}
    \begin{center}
    \begin{small}
    \begin{sc}
    \begin{tabular}{lccc}
        \toprule
        Functionality & CE & TMA & Reg \\
        \midrule
        P2P Transfer & \checkmark & \checkmark & \checkmark \\
        In-fabric Broadcast & \checkmark & \checkmark & \checkmark \\
        P2P Reduction & $\times$ & \checkmark & \checkmark \\
        In-fabric Reduction & $\times$ & $\times$ & \checkmark \\
        Elementwise Transfer & $\times$ & $\times$ & \checkmark \\
        \bottomrule
    \end{tabular}
    \end{sc}
    \end{small}
    \end{center}
    \vspace{-2em}
\end{wraptable}

We find that these mechanisms excel in different scenarios. As illustrated in Figure~\ref{fig:nvl-saturation}, register-level operations require $3.2$--$5.1\times$ more SMs than TMA to saturate NVLink bandwidth, leaving little opportunity for intra-SM overlap. Register instructions are therefore useful when neither the copy engine nor TMA provides the required functionality. A representative case is NVSwitch in-network reduction (e.g., \texttt{multimem.ld\_reduce} and \texttt{multimem.red}), which can substantially speed up workloads like all-reduce. Existing communication libraries do not exploit this design space; for instance, NVSHMEM relies exclusively on register-level operations for intra-node data transfers. Table~\ref{table:comm-method-comparisons} summarizes the functionalities supported by each mechanism.

\subsubsection{Scheduling}
\label{sec:scheduling}

We now examine workload scheduling strategies for multi-GPU kernels. There are two main ways to overlap compute and communication within a kernel:

\begin{enumerate}[itemsep=0.1pt,topsep=0pt,leftmargin=*]
    \item \textit{Intra-SM} overlapping partitions the threads within an SM into two pools: one issuing compute/memory instructions and the other issuing communication instructions. 
    \item \textit{Inter-SM} overlapping partitions the SMs into two pools: one for computation and the other for communication.
\end{enumerate}

While prior work primarily uses inter-SM overlapping~\cite{25-nanoflow,25-comet}, we find that each provides different compute-communication trade-offs depending on workload characteristics.\footnote{The focus on inter-SM overlap in prior work is largely due to pre-Hopper architecture limitations, which lacked single-thread bulk asynchronous transfers and therefore required entire warps or thread blocks to participate, increasing register pressure.}

\paragraph{Intra-SM overlapping.} Intra-SM overlapping is effective when the ideal communication pattern aligns with that of computation, allowing communication to be naturally embedded within the computation pipeline. In such cases, it is superior to inter-SM overlapping for two main reasons:

\begin{enumerate}[itemsep=0.1pt,topsep=0pt,leftmargin=*]
    \item Unlike in inter-SM overlapping, all compute units (i.e., tensor cores) across all SMs are busy in an intra-SM overlapping scheme. This is crucial because, unlike communication bandwidth, compute throughput scales linearly with the number of SMs that perform computation.
    \item Inter-SM communication incurs additional synchronization overhead \( T_{\text{sync}} \), as it must traverse the HBM. Our microbenchmarks show that a single intra-SM synchronization using \texttt{mbarrier} objects incurs approximately 64~ns of latency, whereas inter-SM synchronization through the HBM takes about 832~ns.
\end{enumerate}

\begin{wrapfigure}{r}{0.5\linewidth}
    \vspace{-1em}
    \centering
    \includegraphics[
        width=0.5\textwidth
    ]{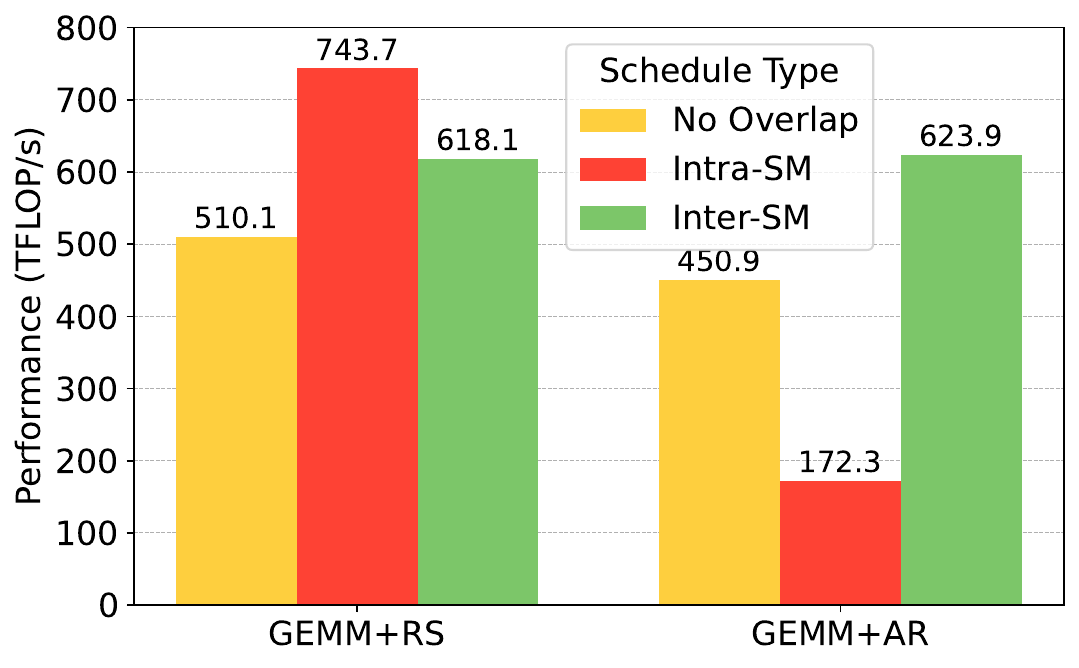}
    \caption{GEMM reduce-scatter (RS) and all-reduce (AR) performance across overlapping schedules. Measured on $8\times$H100 GPUs with local GEMM shape $N \times N \times N/8$ ($N = 32768$) and element type BF16.}
    \label{fig:fused-gemm-diff-schedules}
\end{wrapfigure}

We illustrate these effects using a kernel that fuses a GEMM with a reduce-scatter (RS). Figure~\ref{fig:fused-gemm-diff-schedules} (left) shows that the GEMM+RS kernel achieves higher compute throughput under an intra-SM overlapping schedule, due to higher compute utilization and lower synchronization overhead.

We further show that intra-SM overlapping can almost completely hide communication overhead in certain regimes. Consider an \(M \times N \times K\) GEMM+RS fused kernel with per-iteration tiles of size \(m \times n \times k\). In a typical GEMM kernel, an output tile region is selected, and the \(m \times n \times k\) sub-GEMM is executed \(K / k\) times before the result is stored.

Given the per-element size \(s\), sustained tensor core throughput \(R\) (in FLOP/s), and per-GPU NVLink bandwidth \(B\) (in bytes/s), the compute and communication times for producing a single output tile of size \(m \times n\) are given by:

\[
    T_{\text{comp\_tile}} = \frac{2mnk}{R} \times \frac{K}{k} = \frac{2mnK}{R}
\]

\[
    T_{\text{comm\_tile}} = \frac{smn}{B}
\]

From this, communication can be completely hidden by computation when \(T_{\text{comp\_tile}} \ge T_{\text{comm\_tile}}\), i.e.,

\[
    K \ge \frac{sR}{2B}
\]

For BF16 GEMM on H100 GPUs, \(s = 2\), \(R = 989 \times 10^{12}\), and \(B = 450 \times 10^{9}\), implying that communication is hidden when \(K \gtrsim 2197\). We verify this empirically in Table~\ref{table:intra-overlap-ablation}, where we ablate our fused GEMM+RS kernel against a standalone GEMM kernel. The results show that at \(K = 2048\), the non-overlapped communication ratio drops by roughly half, and beyond that, communication becomes nearly fully hidden. The residual communication time near \(K = 2048\) arises from atomic additions required for output tile accumulation, which prevent complete overlap.

\begin{table}[t]
    \caption{
        Measured BF16 GEMM and GEMM+RS performance (ms).
    }
    \label{table:intra-overlap-ablation}
    \begin{center}
    \begin{small}
    \begin{sc}
    \begin{tabular}{ccccc}
        \toprule
        $M$\&$N$ & $K$ & GEMM & GEMM+RS & Comm Ratio \\
        \midrule
        32768 & 512 & 2.071 & 6.483 & \textbf{68\%} \\
        32768 & 1024 & 2.918 & 6.613 & \textbf{56\%} \\
        32768 & 2048 & 5.567 & 7.531 & \textbf{26\%} \\
        32768 & 4096 & 11.78 & 11.828 & \textbf{$<$1\%} \\
        32768 & 8192 & 23.285 & 25.325 & \textbf{8\%} \\
        \bottomrule
    \end{tabular}
    \end{sc}
    \end{small}
    \end{center}
\end{table}

\paragraph{Inter-SM overlapping.} While intra-SM overlapping fully utilizes GPU compute, it constrains communication to follow the computation pattern. This leads to two potential drawbacks: the inability to exploit in-network acceleration and suboptimal L2 caching behavior. Inter-SM overlapping mitigates these issues but introduces a partitioning trade-off: deciding how many SMs to allocate to communication versus computation.

\textit{In-network acceleration.} Recent networking hardware integrates compute directly into the interconnect fabric, enabling in-network reductions and collective offload within switches and link controllers~\cite{nvswitch,sharp}. This transforms the interconnect from a passive data mover into an active participant in collectives. For communication-heavy kernels such as fused GEMM all-reduce (AR), in-network reduction can significantly reduce bandwidth usage. However, performing it within the same SM is impractical due to register pressure, limited occupancy, and inter-GPU synchronization costs. A more effective approach is to accumulate partial results in HBM, signal completion after each local write, and delegate a few specialized SMs to execute a single in-network all-reduce once all devices finish.

This tradeoff is shown in Figure~\ref{fig:fused-gemm-diff-schedules} (right). Intra-SM overlapping issues \(N\) atomic writes to \(N\) destinations for each output tile, where \(N\) is the number of GPUs. Even with a fully interconnected NVSwitch fabric, each GPU is limited by its 450~GB/s per-port bandwidth, causing concurrent peer writes to serialize at the destination. Inter-SM overlapping reduces \(T_{\text{comm}}\) by roughly a factor of \(N\), typically outweighing the cost of dedicating a few SMs to communication.

\begin{wrapfigure}{r}{0.5\linewidth}
    \vspace{-1.5em}
    \centering
    \includegraphics[
        width=0.5\textwidth
    ]{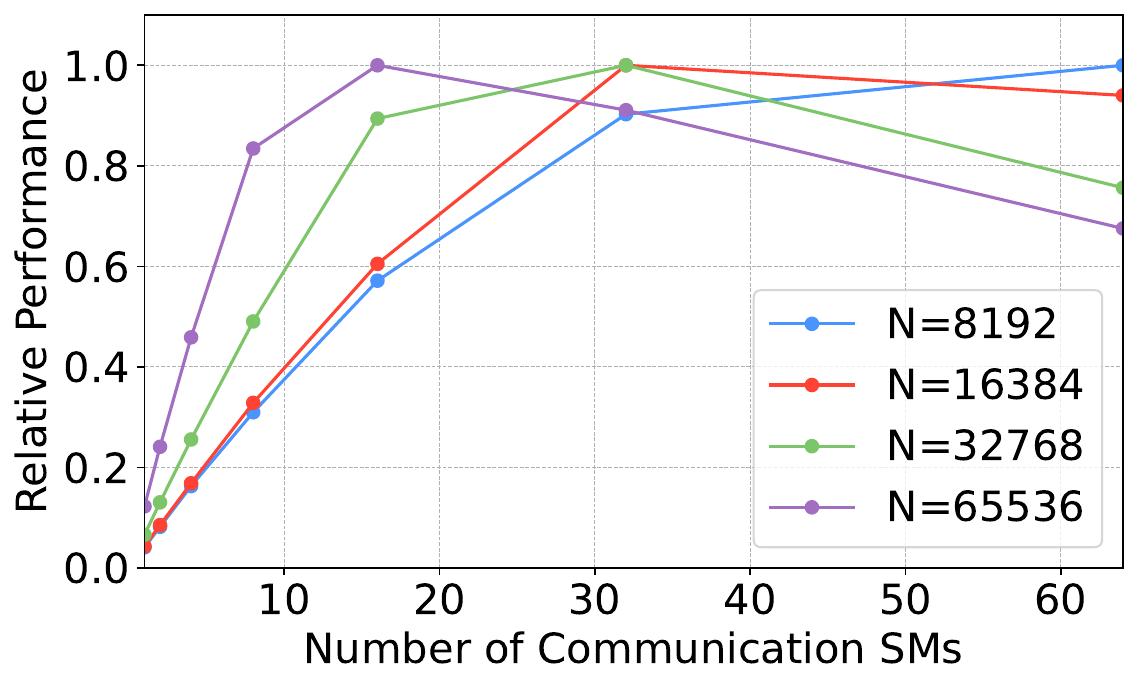}
    \caption{Comparison of different inter-SM scheduling performance on all-gather (AG) GEMM ($N \times N/8 \times N$).}
    \label{fig:inter-sm-schedules}
\end{wrapfigure}

\textit{Remote cache reuse.} Another limitation of intra-SM overlapping arises from the \textit{far-sided} nature of L2 caching for peer HBM accesses. Data fetched from a peer GPU is cached only on the source device, not on the requester. Consequently, every remote access is bottlenecked by NVLink bandwidth. A representative case appears in Ring Attention~\cite{24-ring-attention}, where key and value (KV) tensors are reused across multiple attention blocks. Letting each thread block independently load them from remote GPUs leads to redundant transfers and rapid interconnect saturation. Instead, performing bulk transfers of the next block's \(K\) and \(V\) tensors to local HBM using communication-dedicated SMs, while the remaining SMs compute, substantially reduces \(T_{\text{comm}}\) and improves L2 reuse, as shown in Section~\ref{sec:sequence-parallelism}.

\textit{SM partitioning.} Inter-SM overlapping requires balancing SMs between communication and computation. As shown in Figure~\ref{fig:inter-sm-schedules}, the optimal split depends on input size: larger workloads favor more compute SMs, while smaller ones need proportionally more SMs for communication. \textsc{PK} allows users to automatically search for the optimal SM allocation at runtime through a unified program template.

\subsubsection{Minimizing design overheads}
\label{sec:comm-overheads}

Ideally, abstractions should preserve the developer's ability to achieve peak hardware performance. In practice, however, certain design choices in widely used communication libraries like NCCL and NVSHMEM---particularly in synchronization and buffering---constrain this ability.

\begin{wrapfigure}{r}{0.5\linewidth}
    \vspace{-2em}
    \centering
    \includegraphics[
        width=0.5\textwidth
    ]{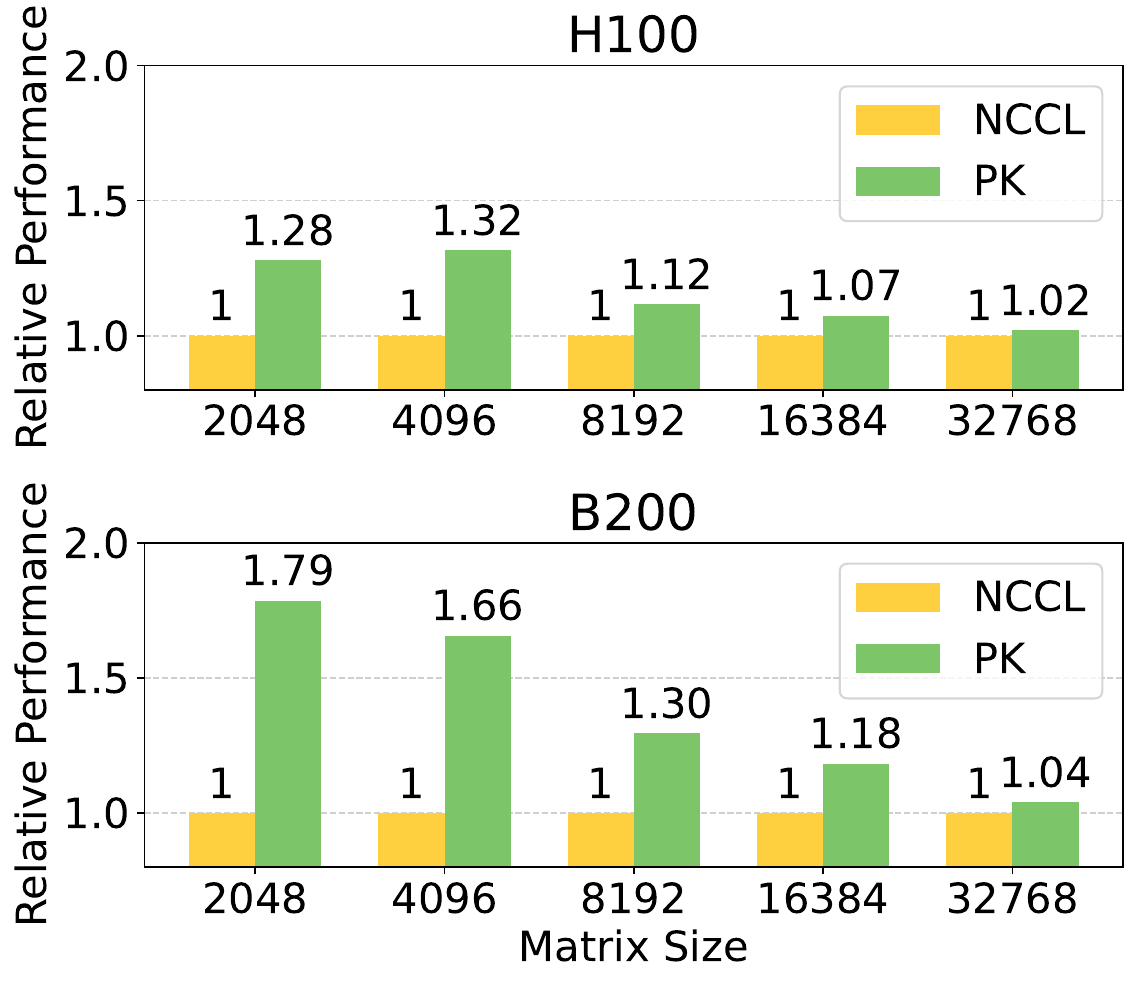}
    \vspace{-2em}
    \caption{All-reduce sum kernel comparison (BF16).}
    \label{fig:all-reduce}
    \vspace{-2em}
\end{wrapfigure}

\paragraph{Two-way synchronization and intermediate buffering.} Many multi-GPU communication libraries impose synchronization and buffering constraints. For example, NCCL enforces two-way synchronization for every operation: both sender and receiver must be ready and acknowledge each other before data transfer begins, even for point-to-point communication. In addition, to reduce peer memory exchange overhead, NCCL employs small pre-allocated intermediate buffers (communication channels), introducing extra data movement. While such overheads are masked for large inputs, they become significant in fine-grained communication. {\shortname} avoids these issues by using pre-allocated destination buffers, enabling direct, one-way transfers without intermediate staging. As shown in Figure~\ref{fig:all-reduce}, this design improves the performance of pure communication kernels such as all-reduce by up to $1.79\times$.

\paragraph{Peer-memory access and synchronization.} NVSHMEM, the de facto low-level standard for multi-GPU communication, also introduces additional overhead in its public API functions. Each remote peer access performs a global memory load (\texttt{\_\_ldg}) to retrieve the peer address and enforces a group synchronization (e.g., \texttt{\_\_syncthreads}). By keeping peer addresses in registers and removing unnecessary synchronizations, \textsc{PK} eliminates these costs, achieving up to $4.5\times$ lower element-wise NVLink access latency and about 20~GB/s higher bandwidth utilization.

\subsection{Abstractions}
\label{sec:abstractions}

We introduce {\name} (\shortname), a collection of abstractions that generalizes the tile-based programming principles proposed in {\tk}---and successor systems such as CuTe DSL and TileLang---to the multi-GPU setting. \shortname\ provides a minimal and complementary set of primitives for efficient multi-GPU communication. These abstractions expose high-performance communication mechanisms, enable the workload scheduling patterns described earlier, and minimize performance overheads by design. {\shortname} hides low-level complexity that does not impact performance, while preserving full user control through its CUDA/C++ embedded design.

\subsubsection{Data Structure}
\label{sec:pk-data-structure}

{\shortname} defines a data structure for each level of the GPU memory hierarchy, as illustrated in Figure~\ref{fig:pull} (left).
At the \textit{register} level, the minimum unit of execution is a \(16\times16\) tile, consistent with the original {\shorttk} design.  
At the \textit{shared memory} level, users operate on shared tiles that enable asynchronous, tile-granularity loads from and stores to peer HBM by a single thread. Store operations optionally support atomic reductions on peer memory and multicast to multiple devices via in-network broadcast. These operations preserve tensor-core–friendly layouts to remain efficient within local compute pipelines.  
At the \textit{HBM} level, {\name} introduces the \textbf{Parallel Global Layout (PGL)}, which represents identically shaped and sized memory regions allocated across all devices. PGL serves as the central data structure enabling asynchronous P2P transfers, broadcasts, and synchronous in-fabric multicasts and reductions over tile-indexed regions.  
All data abstractions enforce essential principles such as coalesced interconnect access, swizzling to minimize bank conflicts and match tensor-core layouts, and fully device-initiated communication.

\subsubsection{Multi-GPU Operations}
\label{sec:pk-primitives}

We introduce eight new primitives, which suffice to implement all kernels demonstrated in Section~\ref{sec:experiments}. The original {\tk} operators are also extended to remain fully compatible with the aforementioned data structures.

\paragraph{P2P communication primitives}

\begin{itemize}[itemsep=0pt,topsep=0pt,leftmargin=*]
    \item \texttt{store\_async(dst, src, coord) // Store a shared tile to multicast memory.}
    \item \texttt{store\_add\_async(dst, src, coord) // Atomically add a shared tile to multicast memory.}
\end{itemize}

\paragraph{Network-accelerated communication primitives}

\begin{itemize}[itemsep=0pt,topsep=0pt,leftmargin=*]
    \item \texttt{reduce(dst, dst\_coord, src, src\_coord) // Reduce data from multicast memory to local HBM.}
    \item \texttt{all\_reduce(dst\_and\_src, coord) // Reduce data from multicast memory and write back to it.}
\end{itemize}

\paragraph{Inter-device and inter-SM synchronization primitives}

\begin{itemize}[itemsep=0pt,topsep=0pt,leftmargin=*]
    \item \texttt{signal(bar, coord, dev\_idx, val) // Signal a specific device's barrier.}
    \item \texttt{signal\_all(bar, coord, val) // Signal all devices' barriers simultaneously.}
    \item \texttt{wait(bar, coord, dev\_idx, expected) // Wait until a device's barrier reaches a value.}
    \item \texttt{barrier(bar, coord, dev\_idx) // Wait for all devices to reach this point.}
\end{itemize}

\vspace{1em}

Because all {\shortname} data structures are tile-based, the new primitives also operate at tile granularity, ranging from \(16\times16\) (the minimum tile) up to the shared-memory limit (about \(256\times256\)). All operations are device-initiated and use coordinates (\texttt{coord}) represented as \texttt{int4} values specifying tile indices in local or remote HBM. P2P primitives are asynchronous and single-threaded, enabling fusion with other operations (e.g., tensor-core compute), whereas network-accelerated primitives require at least warp-level participation for optimal throughput. Synchronization primitives provide simple signaling and waiting mechanisms, enabling users to design arbitrary workload scheduling schemes. A complete API description is provided in Appendix~\ref{appendix-api}.

\subsubsection{Program Template}
\label{sec:pk-program-template}

We provide a unified program template for implementing a wide range of multi-GPU kernels. As shown in Figure~\ref{fig:pull} (right), the template defines four worker components---\textit{loader}, \textit{storer}, \textit{consumer}, and \textit{communicator}---each encapsulating a common warp/SM specialization. The \textit{loader} performs local or peer HBM reads, while the \textit{storer} handles local or peer HBM writes. When either component accesses peer HBM, intra-SM overlapping is employed. The \textit{communicator} occupies one or more SMs exclusively to perform dedicated communication, enabling inter-SM overlapping. Finally, the \textit{consumer} issues tensor- or CUDA-core–based local compute. Beyond providing a structural pattern, the template automates common low-level tasks, including kernel configuration, shared memory and TMA setup, barrier and synchronization management, and tuning for optimal SM/warp partitioning. This allows users to focus solely on the per-tile compute and communication logic. A detailed description of the template and an example kernel are provided in Appendix~\ref{appendix-template-kernels}.

\subsubsection{Utilities}
\label{sec:pk-utilities}

We provide inter-process communication (IPC) and PyTorch utilities for seamless integration with multi-process execution (e.g., via \texttt{torchrun}). These utilities manage low-level OS driver interactions and support pre-allocation of multi-GPU memory, enabling direct P2P communication without intermediate staging overheads. Appendices~\ref{appendix-multi-gpu-setup-process} and \ref{appendix-in-network-accel-setup-process} provide further implementation details.

\section{Experiments}
\label{sec:experiments}

\begin{figure}[t]
    \centering
    \includegraphics[
        width=1.0\textwidth
    ]{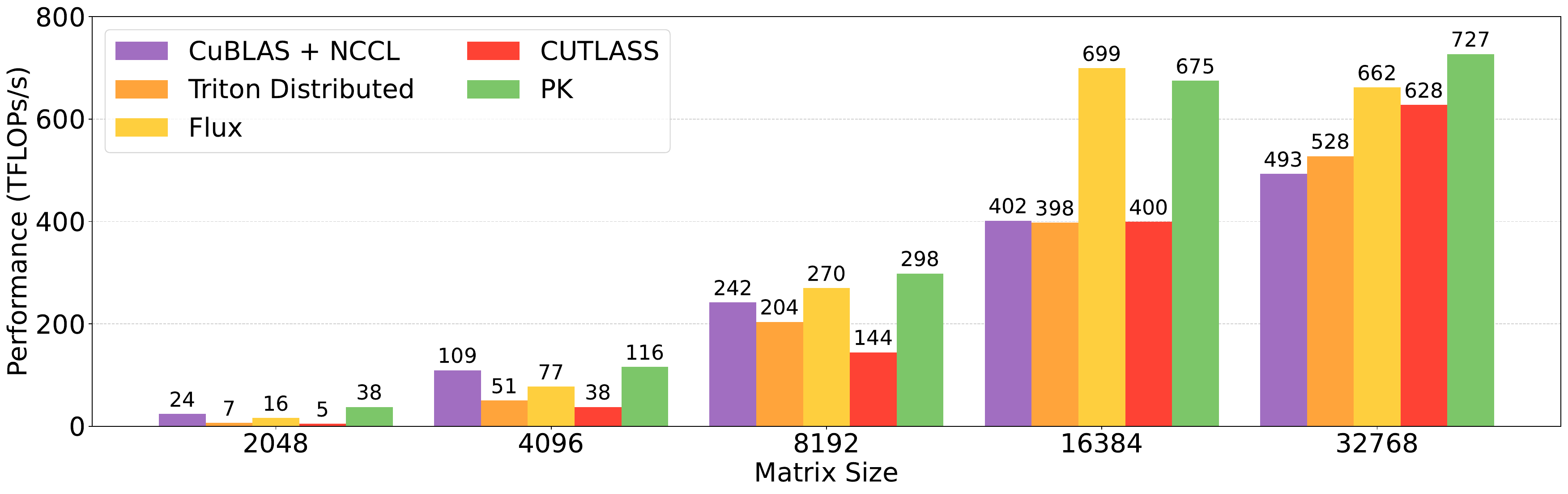}
    \caption{AG + GEMM performance. Local GEMM size is $N \times N/8 \times N$, with $N$ given in the X-axis.}
    \label{fig:ag-gemm-h100}
\end{figure}

\begin{figure}[t]
    \centering
    \includegraphics[
        width=1.0\textwidth
    ]{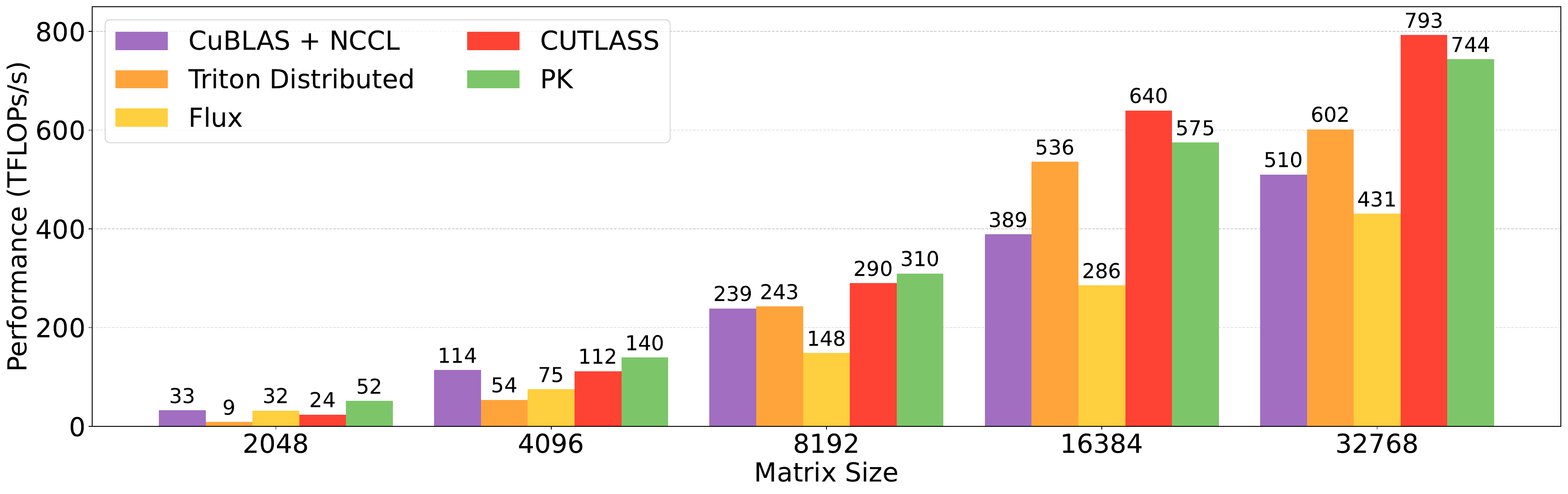}
    \caption{GEMM + RS performance. Local GEMM size is $N \times N \times N/8$, with $N$ given in the X-axis.}
    \label{fig:gemm-rs-h100}
\end{figure}

\begin{figure}[t]
    \centering
    \includegraphics[
        width=0.5\textwidth
    ]{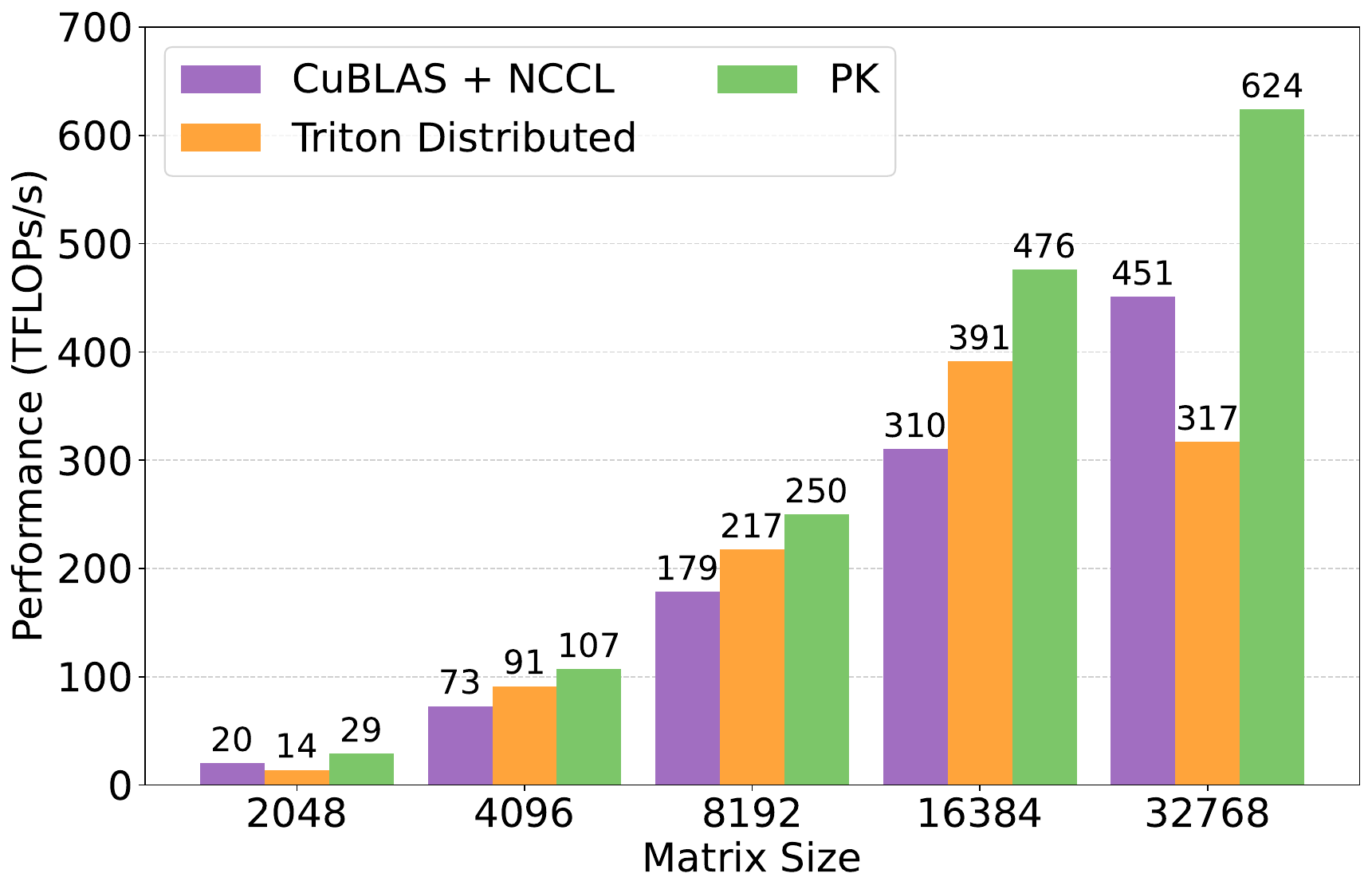}
    \caption{GEMM + AR performance. Local GEMM size is $N \times N \times N/8$, with $N$ given in the X-axis.}
    \label{fig:gemm-ar-h100}
\end{figure}

We demonstrate that {\shortname} generalizes across a diverse range of multi-GPU AI workloads by implementing representative kernels with its abstractions and comparing them against existing frameworks and hand-optimized baselines.

All experiments were conducted using \(8\times\)Nvidia H100 80GB SXM GPUs, interconnected via 4th-generation NVLink and NVSwitch, using CUDA~12.6 and PyTorch~2.8.0. All matrix multiplications use BF16 as the element type and FP32 as the tensor core accumulator type. For brevity, we denote the GEMM shape as $M \times N \times K$, where the first operand has dimensions $M \times K$ and the second has dimensions $K \times N$. We report the observed average compute throughput.

Although the experiments in this section use H100 GPUs, {\shortname} is fully compatible with B200 GPUs and exhibits similar performance characteristics. We present results on Blackwell GPUs in Appendices~\ref{appendix-blackwell-perf} and \ref{appendix-additional-collective-perf}.

\subsection{Data and Tensor Parallelism}
\label{sec:tensor-parallelism}

To efficiently scale large models, weights are often sharded across multiple devices using \textit{tensor parallelism}~\cite{19-megatron, 22-alpa}, which partitions weight matrices along the row or column dimension. A common strategy combines this with \textit{data parallelism}~\cite{21-efficiently}: inputs sharded by rows are first all-gathered (AG), followed by a GEMM with column-sharded weights, a non-linear activation, and a second GEMM with row-sharded weights, after which a reduce-scatter (RS) or all-reduce (AR) is applied. Communication and computation are overlapped by pairing AG with the first GEMM (AG+GEMM) and RS or AR with the second (GEMM+RS, GEMM+AR).

For these workloads, we compare against the cuBLAS GEMM combined with NCCL as the non-overlapped baseline, compiler-based approaches (Triton Distributed), and hand-optimized kernels (Flux and CUTLASS). Flux and CUTLASS do not provide GEMM–AR kernels and are therefore omitted in those cases. Figures~\ref{fig:ag-gemm-h100}, \ref{fig:gemm-rs-h100}, and \ref{fig:gemm-ar-h100} show the results. Overall, {\shortname} achieves a $1.06$--$1.68\times$ speedup over the non-overlapped baseline and outperforms compiler-based approaches by $1.07$--$5.63\times$. Compared to hand-optimized kernels, {\shortname} matches or surpasses their performance, achieving $0.97$--$2.33\times$ speedup over Flux and $0.90$--$7.39\times$ over CUTLASS. We also note that AG+GEMM and GEMM+RS are often used back-to-back in practice, and no single baseline outperforms {\shortname} when both are combined.

We further observe that compiler-based approaches can exhibit inconsistent performance across diverse hardware platforms. For instance, Triton Distributed, originally developed for H800 GPUs, sometimes performs below the non-overlapped baseline on H100s. Hand-tuned kernels also show reduced efficiency on certain problem shapes.

Under sufficiently large reduction axes, the non-overlapped portion of communication time in {\shortname} falls below 1\%. The communication component of our kernels (excluding GEMM) is implemented in fewer than 50 lines of device code, using the primitives introduced in Section~\ref{sec:abstractions}.

\subsection{Sequence Parallelism}
\label{sec:sequence-parallelism}

\begin{figure}[b]
    \centering
    \includegraphics[
        width=0.48\textwidth
    ]{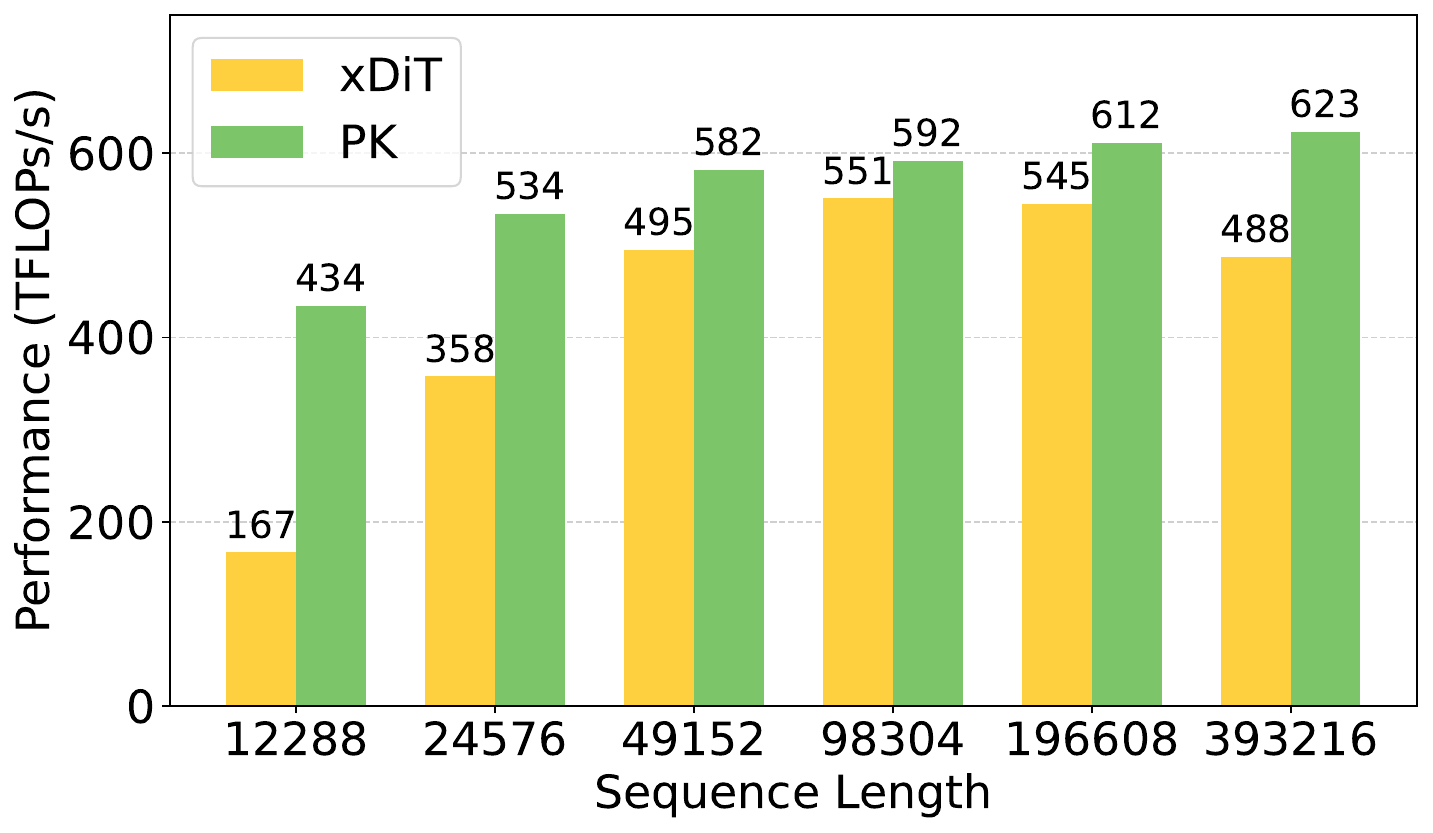}
    \caption{Ring Attention performance across sequence lengths ($B=16$, $H=16$, $D=128$).}
    \label{fig:ring-attention-h100}
\end{figure}

Modern AI workloads increasingly involve inputs with long sequence lengths, requiring a single sequence to be distributed across multiple devices. While sharding along the sequence dimension has minimal impact on MLP or MoE layers, attention layers require each token to attend to all others within the same sequence. This necessitates sequence-parallel approaches such as Ring Attention~\cite{24-ring-attention} and DeepSpeed-Ulysses~\cite{24-ulysses}. In our evaluation, we compare against the state-of-the-art implementations: xDiT~\cite{24-xdit} for Ring Attention and YunChang~\cite{24-yunchang} for DeepSpeed-Ulysses.

\paragraph{Ring Attention.} In Ring Attention, key–value (KV) tensors are partitioned across devices, with each GPU computing blockwise attention on its local shard while concurrently transmitting it to a peer. The baseline xDiT implementation overlaps computation and KV exchange coarsely by launching NCCL P2P sends and FlashAttention-3 kernels on separate CUDA streams. In contrast, {\shortname} can fuse these into a single kernel with explicit inter-SM overlap, precisely allocating SMs between computation and communication, deciding how they synchronize, and auto-tuning this partitioning for optimal performance. As shown in Figure~\ref{fig:ring-attention-h100}, this yields a $1.07\times$–$4.08\times$ speedup over the baseline---evaluated at total sequence lengths (shown on the X-axis)\footnote{Sequence lengths are intentionally set as multiples of 768 because this is required by the original {\shorttk} attention forward kernel.} evenly partitioned across 8 devices---and reduces the non-overlapped communication fraction down to 9\%. 

\begin{figure}[t]
    \centering
    \includegraphics[
        width=0.5\textwidth
    ]{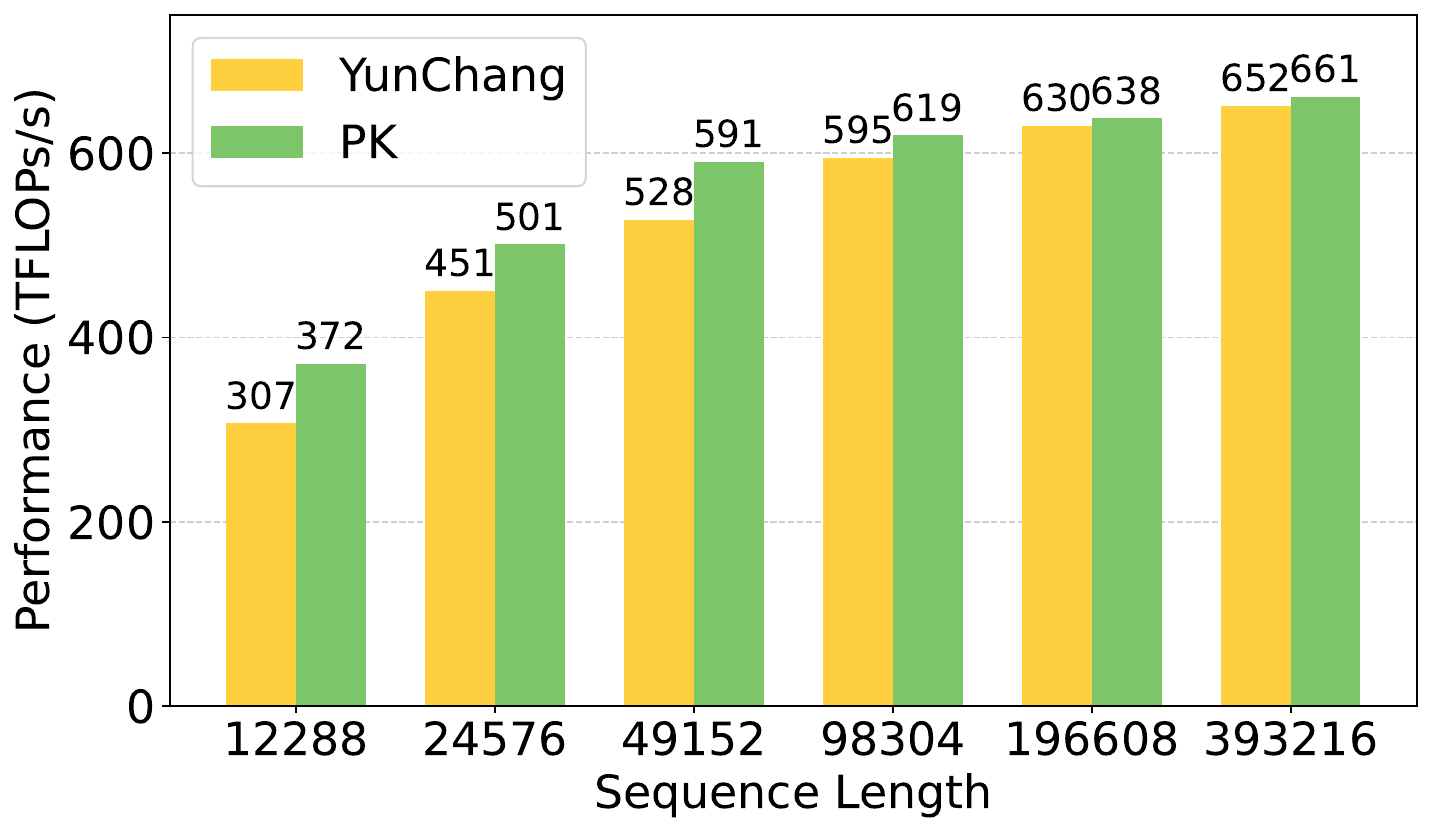}
    \caption{DeepSpeed-Ulysses attention layer performance across sequence lengths ($B=16$, $H=128$, $D=128$).}
    \label{fig:ulysses-attention-h100}
\end{figure}

\begin{figure}[t]
    \centering
    \includegraphics[
        width=0.5\textwidth
    ]{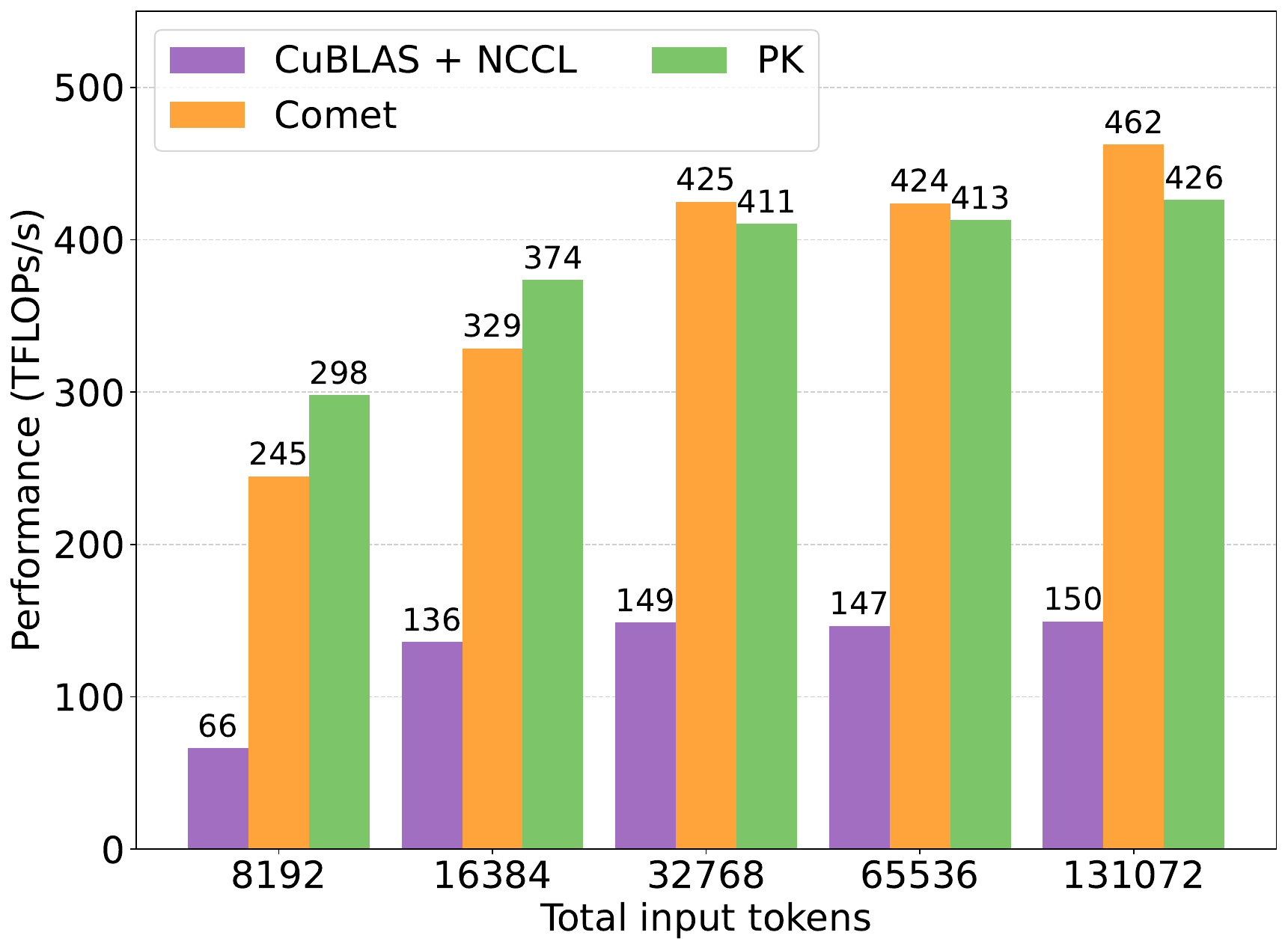}
    \caption{Expert-parallel token dispatch + GEMM performance ($\text{TopK}=8$, $N_{\text{experts}}=256$, $H=7168$, $H_{\text{expert}}=2048$).}
    \label{fig:comet-ep-h100}
\end{figure}

\paragraph{DeepSpeed-Ulysses.} In DeepSpeed-Ulysses, an all-to-all exchange occurs before and after self-attention. Everything except self-attention is sequence-sharded, while self-attention remains head-sharded. The main bottleneck is the fine-grained all-to-all; as NCCL does not natively support this along the inner dimension, the baseline relies on tensor reshaping before and after communication. Using {\shortname}, we implement a fine-grained all-to-all kernel that removes this overhead. As shown in Figure~\ref{fig:ulysses-attention-h100}, this yields a $1.01\times$–$1.39\times$ speedup, evaluated at total sequence lengths (shown on the X-axis) evenly split across 8 devices. The complete kernel remains under 50 lines of device code.

\subsection{Expert Parallelism}
\label{sec:expert-parallelism}

To scale architectures with MoE layers~\cite{17-moe}, multiple experts are distributed evenly across devices, a strategy known as expert parallelism. However, this approach requires costly scattering and gathering of tokens before and after the expert MLP layers. Several approaches mitigate this by overlapping token communication with GEMM computation~\cite{25-deepep,25-comet,25-flashdmoe}. We compare against \textsc{Comet}~\cite{25-comet}, the state-of-the-art fine-grained overlapping strategy for expert parallelism. For demonstration, we evaluate the first half of the MoE layer: overlapping token dispatch with the first expert MLP. As shown in Figure~\ref{fig:comet-ep-h100}, where the total set of input tokens (shown on the X-axis) is initially partitioned evenly across devices, {\shortname} matches or surpasses the hand-tuned baseline, achieving $0.92$–$1.22\times$ the performance of Comet, with fewer than 40 lines of device code added to a grouped GEMM kernel.

\section{Conclusion}
\label{sec:conclusion}

This work presents {\name}, a minimal and systematic framework for building high-performance multi-GPU kernels. By formalizing the design space through three key principles---transfer mechanisms, scheduling strategies, and design overheads---we demonstrate that a small set of primitives can match or surpass the performance of hand-optimized kernels while greatly simplifying implementation. As this work focuses on intra-node execution, extending these abstractions to inter-node communication remains an important direction for future work. At the same time, intra-node systems are rapidly scaling, as shown by Nvidia’s NVL72 and upcoming NVL144, NVL576 architectures, which makes the study of efficient intra-node kernel design increasingly critical for distributed AI workloads.

Our framework and kernels are open sourced at: \url{https://github.com/HazyResearch/ThunderKittens}.

\section*{Acknowledgements}

We are grateful to Cursor and Together AI for making this work possible. 
We thank Dylan Lim for his assistance with the initial implementation of PGL operations.
We thank Yasa Baig, Kelly Buchanan, Francois Chaubard, Mayee Chen, Catherine Deng, Andy Dimnaku, Owen Dugan, Daniel Y. Fu, Roberto Garcia, Ronny Junkins, Ishane Khare, Hermann Kumbong, Jerry Liu, Avanika Narayan, Jon Saad-Falcon, and Alex Waitz for helpful feedback and discussions during this work.
We gratefully acknowledge the support of NIH under No. U54EB020405 (Mobilize), NSF under Nos. CCF2247015 (Hardware-Aware), CCF1763315 (Beyond Sparsity), CCF1563078 (Volume to Velocity), and 1937301 (RTML); US DEVCOM ARL under Nos. W911NF-23-2-0184 (Long-context) and W911NF-21-2-0251 (Interactive Human-AI Teaming); ONR under Nos. N000142312633 (Deep Signal Processing); Stanford HAI under No. 247183; NXP, Xilinx, LETI-CEA, Intel, IBM, Microsoft, NEC, Toshiba, TSMC, ARM, Hitachi, BASF, Accenture, Ericsson, Qualcomm, Analog Devices, Google Cloud, Salesforce, Total, the HAI-GCP Cloud Credits for Research program,  the Stanford Data Science Initiative (SDSI), and members of the Stanford DAWN project: Meta, Google, and VMWare. The U.S. Government is authorized to reproduce and distribute reprints for Governmental purposes notwithstanding any copyright notation thereon. Any opinions, findings, and conclusions or recommendations expressed in this material are those of the authors and do not necessarily reflect the views, policies, or endorsements, either expressed or implied, of NIH, ONR, or the U.S. Government.

\bibliographystyle{references}
\bibliography{references}

\begin{thebibliography}{37}
\providecommand{\natexlab}[1]{#1}
\providecommand{\url}[1]{\texttt{#1}}
\expandafter\ifx\csname urlstyle\endcsname\relax
  \providecommand{\doi}[1]{doi: #1}\else
  \providecommand{\doi}{doi: \begingroup \urlstyle{rm}\Url}\fi

\bibitem[Aimuyo et~al.(2025)Aimuyo, Oh, and Singh]{25-flashdmoe}
Osayamen~Jonathan Aimuyo, Byungsoo Oh, and Rachee Singh.
\newblock {FlashDMoE: Fast Distributed MoE in a Single Kernel}.
\newblock \emph{arXiv preprint arXiv:2506.04667}, June 2025.

\bibitem[AMD(2025)]{25-iris}
AMD.
\newblock Iris: First-class multi-gpu programming experience in triton.
\newblock \url{https://github.com/ROCm/iris}, 2025.

\bibitem[Chang et~al.(2024)Chang, Bao, Hou, Jiang, Zheng, Zhang, Song, Jiang, Lin, and Liu]{24-flux}
Liwen Chang, Wenlei Bao, Qi~Hou, Chengquan Jiang, Ningxin Zheng, Xuanrun Zhang, Zuquan Song, Ziheng Jiang, Haibin Lin, and Xin Liu.
\newblock {FLUX: Fast Software-based Communication Overlap On GPUs Through Kernel Fusion}.
\newblock \emph{arXiv preprint arXiv:2406.06858v1}, June 2024.

\bibitem[Dao et~al.(2022)Dao, Fu, Ermon, Rudra, and R{\'e}]{22-FA1}
Tri Dao, Daniel~Y. Fu, Stefano Ermon, Atri Rudra, and Christopher R{\'e}.
\newblock Flash{A}ttention: Fast and memory-efficient exact attention with {IO}-awareness.
\newblock In \emph{Advances in Neural Information Processing Systems (NeurIPS)}, 2022.

\bibitem[DeepSeek-AI et~al.(2024)DeepSeek-AI, Liu, Feng, Xue, Wang, Wu, Lu, Zhao, Deng, Zhang, Ruan, Dai, Guo, Yang, Chen, Ji, Li, Lin, Dai, Luo, Hao, Chen, Li, Zhang, Bao, Xu, Wang, Zhang, Ding, Xin, Gao, Li, Qu, Cai, Liang, Guo, Ni, Li, Wang, Chen, Chen, Yuan, Qiu, Li, Song, Dong, Hu, Gao, Guan, Huang, Yu, Wang, Zhang, Xu, Xia, Zhao, Wang, Zhang, Li, Wang, Zhang, Zhang, Tang, Li, Tian, Huang, Wang, Zhang, Wang, Zhu, Chen, Du, Chen, Jin, Ge, Zhang, Pan, Wang, Xu, Zhang, Chen, Li, Lu, Zhou, Chen, Wu, Ye, Ma, Wang, Zhou, Yu, Zhou, Pan, Wang, Yun, Pei, Sun, Xiao, Zeng, Zhao, An, Liu, Liang, Gao, Yu, Zhang, Li, Jin, Wang, Bi, Liu, Wang, Shen, Chen, Zhang, Chen, Nie, Sun, Wang, Cheng, Liu, Xie, Liu, Yu, Song, Shan, Zhou, Yang, Li, Su, Lin, Li, Wang, Wei, Zhu, Zhang, Xu, Huang, Li, Zhao, Sun, Li, Wang, Yu, Zheng, Zhang, Shi, Xiong, He, Tang, Piao, Wang, Tan, Ma, Liu, Guo, Wu, Ou, Zhu, Wang, Gong, Zou, He, Zha, Xiong, Ma, Yan, Luo, You, Liu, Zhou, Wu, Ren, Ren, Sha, Fu, Xu, Huang, Zhang, Xie, Zhang, Hao, Gou, Ma,
  Yan, Shao, Xu, Wu, Zhang, Li, Gu, Zhu, Liu, Li, Xie, Song, Gao, and Pan]{24-dsv3}
DeepSeek-AI, Aixin Liu, Bei Feng, Bing Xue, Bingxuan Wang, Bochao Wu, Chengda Lu, Chenggang Zhao, Chengqi Deng, Chenyu Zhang, Chong Ruan, Damai Dai, Daya Guo, Dejian Yang, Deli Chen, Dongjie Ji, Erhang Li, Fangyun Lin, Fucong Dai, Fuli Luo, Guangbo Hao, Guanting Chen, Guowei Li, H.~Zhang, Han Bao, Hanwei Xu, Haocheng Wang, Haowei Zhang, Honghui Ding, Huajian Xin, Huazuo Gao, Hui Li, Hui Qu, J.L. Cai, Jian Liang, Jianzhong Guo, Jiaqi Ni, Jiashi Li, Jiawei Wang, Jin Chen, Jingchang Chen, Jingyang Yuan, Junjie Qiu, Junlong Li, Junxiao Song, Kai Dong, Kai Hu, Kaige Gao, Kang Guan, Kexin Huang, Kuai Yu, Lean Wang, Lecong Zhang, Lei Xu, Leyi Xia, Liang Zhao, Litong Wang, Liyue Zhang, Meng Li, Miaojun Wang, Mingchuan Zhang, Minghua Zhang, Minghui Tang, Mingming Li, Ning Tian, Panpan Huang, Peiyi Wang, Peng Zhang, Qiancheng Wang, Qihao Zhu, Qinyu Chen, Qiushi Du, R.J. Chen, R.L. Jin, Ruiqi Ge, Ruisong Zhang, Ruizhe Pan, Runji Wang, Runxin Xu, Ruoyu Zhang, Ruyi Chen, S.S. Li, Shanghao Lu, Shangyan Zhou, Shanhuang
  Chen, Shaoqing Wu, Shengfeng Ye, Shirong Ma, Shiyu Wang, Shuang Zhou, Shuiping Yu, Shunfeng Zhou, Shuting Pan, T.~Wang, Tao Yun, Tian Pei, Tianyu Sun, W.L. Xiao, Wangding Zeng, Wanjia Zhao, Wei An, Wen Liu, Wenfeng Liang, Wenjun Gao, Wenqin Yu, Wentao Zhang, X.Q. Li, Xiangyue Jin, Xianzu Wang, Xiao Bi, Xiaodong Liu, Xiaohan Wang, Xiaojin Shen, Xiaokang Chen, Xiaokang Zhang, Xiaosha Chen, Xiaotao Nie, Xiaowen Sun, Xiaoxiang Wang, Xin Cheng, Xin Liu, Xin Xie, Xingchao Liu, Xingkai Yu, Xinnan Song, Xinxia Shan, Xinyi Zhou, Xinyu Yang, Xinyuan Li, Xuecheng Su, Xuheng Lin, Y.K. Li, Y.Q. Wang, Y.X. Wei, Y.X. Zhu, Yang Zhang, Yanhong Xu, Yanping Huang, Yao Li, Yao Zhao, Yaofeng Sun, Yaohui Li, Yaohui Wang, Yi~Yu, Yi~Zheng, Yichao Zhang, Yifan Shi, Yiliang Xiong, Ying He, Ying Tang, Yishi Piao, Yisong Wang, Yixuan Tan, Yiyang Ma, Yiyuan Liu, Yongqiang Guo, Yu~Wu, Yuan Ou, Yuchen Zhu, Yuduan Wang, Yue Gong, Yuheng Zou, Yujia He, Yukun Zha, Yunfan Xiong, Yunxian Ma, Yuting Yan, Yuxiang Luo, Yuxiang You, Yuxuan Liu,
  Yuyang Zhou, Z.F. Wu, Z.Z. Ren, Zehui Ren, Zhangli Sha, Zhe Fu, Zhean Xu, Zhen Huang, Zhen Zhang, Zhenda Xie, Zhengyan Zhang, Zhewen Hao, Zhibin Gou, Zhicheng Ma, Zhigang Yan, Zhihong Shao, Zhipeng Xu, Zhiyu Wu, Zhongyu Zhang, Zhuoshu Li, Zihui Gu, Zijia Zhu, Zijun Liu, Zilin Li, Ziwei Xie, Ziyang Song, Ziyi Gao, and Zizheng Pan.
\newblock {DeepSeek-V3 Technical Report}.
\newblock \emph{arXiv preprint arXiv:2412.19437}, December 2024.

\bibitem[Fang \& Zhao(2024)Fang and Zhao]{24-yunchang}
Jiarui Fang and Shangchun Zhao.
\newblock A unified sequence parallelism approach for long context generative ai.
\newblock \emph{arXiv preprint arXiv:2405.07719}, 2024.

\bibitem[Fang et~al.(2024)Fang, Pan, Sun, Li, and Wang]{24-xdit}
Jiarui Fang, Jinzhe Pan, Xibo Sun, Aoyu Li, and Jiannan Wang.
\newblock xdit: an inference engine for diffusion transformers (dits) with massive parallelism.
\newblock \emph{arXiv preprint arXiv:2411.01738}, 2024.

\bibitem[Google(2025)]{25-pallas}
Google.
\newblock Pallas: a jax kernel language, 2025.
\newblock URL \url{https://docs.jax.dev/en/latest/pallas/index.html}.

\bibitem[He et~al.(2024)He, Wright, Wehrstedt, Liu, and Liang]{24-tp-async}
Horace He, Less Wright, Luca Wehrstedt, Tianyu Liu, and Wanchao Liang.
\newblock Introducing async tensor parallelism in pytorch.
\newblock https://discuss.pytorch.org/t/distributed-w-torchtitan-introducing-async-tensor-parallelism-in-pytorch/209487/1, September 2024.

\bibitem[Jacobs et~al.(2024)Jacobs, Tanaka, Zhang, Zhang, Song, Rajbhandari, and He]{24-ulysses}
Sam~Ade Jacobs, Masahiro Tanaka, Chengming Zhang, Minjia Zhang, Shuaiwen~Leon Song, Samyam Rajbhandari, and Yuxiong He.
\newblock System optimizations for enabling training of extreme long sequence transformer models.
\newblock In \emph{Proceedings of the 43rd ACM Symposium on Principles of Distributed Computing (PODC '24)}, pp.\  121--130, New York, NY, USA, 2024. Association for Computing Machinery.
\newblock \doi{10.1145/3662158.3662806}.
\newblock URL \url{https://doi.org/10.1145/3662158.3662806}.

\bibitem[Jia et~al.(2019)Jia, Zaharia, and Aiken]{19-flexflow}
Zhihao Jia, Matei Zaharia, and Alex Aiken.
\newblock {Beyond Data and Model Parallelism for Deep Neural Networks}.
\newblock \emph{Proceedings of the 2nd SysML Conference}, 2019.

\bibitem[Liang et~al.(2025)Liang, Liu, Wright, Constable, Gu, Huang, Zhang, Feng, Huang, Wang, Purandare, Nadathur, and Idreos]{25-torchtitan}
Wanchao Liang, Tianyu Liu, Less Wright, Will Constable, Andrew Gu, Chien-Chin Huang, Iris Zhang, Wei Feng, Howard Huang, Junjie Wang, Sanket Purandare, Gokul Nadathur, and Stratos Idreos.
\newblock Torchtitan: One-stop pytorch native solution for production ready {LLM} pretraining.
\newblock In \emph{The Thirteenth International Conference on Learning Representations}, 2025.
\newblock URL \url{https://openreview.net/forum?id=SFN6Wm7YBI}.

\bibitem[Liu et~al.(2024)Liu, Zaharia, and Abbeel]{24-ring-attention}
Hao Liu, Matei Zaharia, and Pieter Abbeel.
\newblock Ringattention with blockwise transformers for near-infinite context.
\newblock In \emph{The Twelfth International Conference on Learning Representations}, 2024.
\newblock URL \url{https://openreview.net/forum?id=WsRHpHH4s0}.

\bibitem[Narayanan et~al.(2021)Narayanan, Shoeybi, Casper, LeGresley, Patwary, Korthikanti, Vainbrand, Kashinkunti, Bernauer, Catanzaro, Phanishayee, and Zaharia]{21-efficiently}
Deepak Narayanan, Mohammad Shoeybi, Jared Casper, Patrick LeGresley, Mostofa Patwary, Vijay Korthikanti, Dmitri Vainbrand, Prethvi Kashinkunti, Julie Bernauer, Bryan Catanzaro, Amar Phanishayee, and Matei Zaharia.
\newblock {Efficient Large-Scale Language Model Training on GPU Clusters Using Megatron-LM}.
\newblock \emph{Proceedings of the International Conference for High Performance Computing, Networking, Storage and Analysis}, 2021.

\bibitem[{Nvidia}()]{nvlink-nvswitch}
{Nvidia}.
\newblock {Nvidia NVLink and NVLink Switch}.
\newblock \url{https://www.nvidia.com/en-us/data-center/nvlink/}.

\bibitem[Nvidia()]{nvshmem}
Nvidia.
\newblock Nvshmem.
\newblock \url{https://developer.nvidia.com/nvshmem}.

\bibitem[Nvidia(2020)]{20-ampere}
Nvidia.
\newblock Nvidia ampere architecture in-depth.
\newblock \url{https://developer.nvidia.com/blog/nvidia-ampere-architecture-in-depth/}, May 2020.

\bibitem[{Nvidia}(2024)]{24-cute-dsl}
{Nvidia}.
\newblock {Nvidia CuTe}.
\newblock \url{https://github.com/NVIDIA/cutlass/blob/main/media/docs/cute/00_quickstart.md}, 2024.

\bibitem[Nvidia(2024)]{sharp}
Nvidia.
\newblock Advancing performance with nvidia sharp in-network computing.
\newblock \url{https://developer.nvidia.com/blog/advancing-performance-with-nvidia-sharp-in-network-computing/}, 2024.

\bibitem[Nvidia(2025)]{25-blackwell}
Nvidia.
\newblock Nvidia blackwell architecture technical brief.
\newblock \url{https://resources.nvidia.com/en-us-blackwell-architecture}, 2025.

\bibitem[{Nvidia}(2025)]{25-nvl-roadmap}
{Nvidia}.
\newblock {Company Overview}.
\newblock \url{https://s201.q4cdn.com/141608511/files/doc_presentations/2025/08/Q226-NVDA-Company-Overview-Final.pdf}, August 2025.

\bibitem[Nvidia(2025{\natexlab{a}})]{nccl}
Nvidia.
\newblock Nvidia collective communications library (nccl).
\newblock \url{https://developer.nvidia.com/nccl}, 2025{\natexlab{a}}.

\bibitem[Nvidia(2025{\natexlab{b}})]{nvswitch}
Nvidia.
\newblock Nvidia nvlink and nvlink switch.
\newblock \url{https://www.nvidia.com/en-us/data-center/nvlink/}, 2025{\natexlab{b}}.

\bibitem[Shazeer et~al.(2017)Shazeer, Mirhoseini, Maziarz, Davis, Le, Hinton, and Dean]{17-moe}
Noam Shazeer, Azalia Mirhoseini, Krzysztof Maziarz, Andy Davis, Quoc Le, Geoffrey Hinton, and Jeff Dean.
\newblock Outrageously large neural networks: The sparsely-gated mixture-of-experts layer.
\newblock In \emph{International Conference on Learning Representations}, 2017.
\newblock \doi{1701.06538}.
\newblock URL \url{https://arxiv.org/abs/1701.06538}.

\bibitem[Shoeybi et~al.(2019)Shoeybi, Patwary, Puri, LeGresley, Casper, and Catanzaro]{19-megatron}
Mohammad Shoeybi, Mostofa Patwary, Raul Puri, Patrick LeGresley, Jared Casper, and Bryan Catanzaro.
\newblock {Megatron-LM: Training Multi-Billion Parameter Language Models Using Model Parallelism}.
\newblock \emph{arXiv preprint arXiv:1909.08053}, September 2019.

\bibitem[Si et~al.(2025)Si, Balaji, Chen, Chu, Gangidi, Hasan, Iyengar, Johnson, Liu, Ren, Shetty, Steinbrecher, Wang, Wu, Xie, Yang, Yang, Yu, Yu, Zhao, Bland, Boyda, Gumudavelli, Kannan, Lumezanu, Miao, Qu, Ramesh, Samoylov, Seidel, Sundaresan, Tian, Tan, Zhang, Zhao, Zheng, Zhu, and Zeng]{25-ncclx}
Min Si, Pavan Balaji, Yongzhou Chen, Ching-Hsiang Chu, Adi Gangidi, Saif Hasan, Subodh Iyengar, Dan Johnson, Bingzhe Liu, Regina Ren, Ashmitha~Jeevaraj Shetty, Greg Steinbrecher, Yulun Wang, Bruce Wu, Xinfeng Xie, Jingyi Yang, Mingran Yang, Kenny Yu, Minlan Yu, Cen Zhao, Wes Bland, Denis Boyda, Suman Gumudavelli, Prashanth Kannan, Cristian Lumezanu, Rui Miao, Zhe Qu, Venkat Ramesh, Maxim Samoylov, Jan Seidel, Srikanth Sundaresan, Feng Tian, Qiye Tan, Shuqiang Zhang, Yimeng Zhao, Shengbao Zheng, Art Zhu, and Hongyi Zeng.
\newblock Collective communication for 100k+ gpus.
\newblock \emph{arXiv preprint arXiv:2510.20171}, October 2025.

\bibitem[Spector et~al.(2025)Spector, Arora, Singhal, Parthasarathy, Fu, and Ré]{25-thunderkittens}
Benjamin~F. Spector, Simran Arora, Aaryan Singhal, Arjun Parthasarathy, Daniel~Y. Fu, and Christopher Ré.
\newblock Thunderkittens: Simple, fast, and adorable kernels.
\newblock In \emph{The Thirteenth International Conference on Learning Representations}, April 2025.
\newblock URL \url{https://openreview.net/forum?id=0fJfVOSUra}.

\bibitem[Thakkar et~al.()Thakkar, Ramani, Cecka, Shivam, Lu, Yan, Kosaian, Hoemmen, Wu, Kerr, Nicely, Merrill, Blasig, Qiao, Majcher, Springer, Hohnerbach, Wang, and Gupta]{cutlass}
Vijay Thakkar, Pradeep Ramani, Cris Cecka, Aniket Shivam, Honghao Lu, Ethan Yan, Jack Kosaian, Mark Hoemmen, Haicheng Wu, Andrew Kerr, Matt Nicely, Duane Merrill, Dustyn Blasig, Fengqi Qiao, Piotr Majcher, Paul Springer, Markus Hohnerbach, Jin Wang, and Manish Gupta.
\newblock Cutlass: Cuda templates for linear algebra subroutines.
\newblock \url{https://github.com/NVIDIA/cutlass}.

\bibitem[Tillet et~al.(2019)Tillet, Kung, and Cox]{19-triton}
Philippe Tillet, H.~T. Kung, and David Cox.
\newblock Triton: an intermediate language and compiler for tiled neural network computations.
\newblock In \emph{Proceedings of the 3rd ACM SIGPLAN International Workshop on Machine Learning and Programming Languages}, 2019.

\bibitem[Tsu(2022)]{22-hgx-h100}
William Tsu.
\newblock {Introducing Nvidia HGX H100: An Accelerated Server Platform for AI and High-Performance Computing}.
\newblock \url{https://developer.nvidia.com/blog/introducing-nvidia-hgx-h100-an-accelerated-server-platform-for-ai-and-high-performance-computing/}, April 2022.

\bibitem[Zhang et~al.(2025)Zhang, Zheng, Lin, Jiang, Bao, Jiang, Hou, Cui, Zheng, Chang, Chen, and Liu]{25-comet}
Shulai Zhang, Ningxin Zheng, Haibin Lin, Ziheng Jiang, Wenlei Bao, Chengquan Jiang, Qi~Hou, Weihao Cui, Size Zheng, Li-Wen Chang, Quan Chen, and Xin Liu.
\newblock {Comet: Fine-grained Computation-communication Overlapping for Mixture-of-Experts}.
\newblock \emph{Proceedings of the 8th MLSys Conference}, March 2025.

\bibitem[Zhao et~al.(2025)Zhao, Zhou, Zhang, Deng, Xu, Liu, Yu, Li, and Zhao]{25-deepep}
Chenggang Zhao, Shangyan Zhou, Liyue Zhang, Chengqi Deng, Zhean Xu, Yuxuan Liu, Kuai Yu, Jiashi Li, and Liang Zhao.
\newblock Deepep: an efficient expert-parallel communication library.
\newblock \url{https://github.com/deepseek-ai/DeepEP}, 2025.

\bibitem[Zhao et~al.(2023)Zhao, Gu, Varma, Luo, Huang, Xu, Wright, Shojanazeri, Ott, Shleifer, Desmaison, Balioglu, Damania, Nguyen, Chauhan, Hao, Mathews, and Li]{zhao2023fsdp}
Yanli Zhao, Andrew Gu, Rohan Varma, Liang Luo, Chien-Chin Huang, Min Xu, Less Wright, Hamid Shojanazeri, Myle Ott, Sam Shleifer, Alban Desmaison, Can Balioglu, Pritam Damania, Bernard Nguyen, Geeta Chauhan, Yuchen Hao, Ajit Mathews, and Shen Li.
\newblock Pytorch fsdp: Experiences on scaling fully sharded data parallel, 2023.

\bibitem[Zheng et~al.(2022)Zheng, Li, Zhang, Zhuang, Chen, Huang, Wang, Xu, Zhuo, and Xing]{22-alpa}
Lianmin Zheng, Zhuohan Li, Hao~Zhang Zhang, Yonghao Zhuang, Zhifeng Chen, Yanping Huang, Yida Wang, Yuanzhong~Xu Xu, Danyang Zhuo, and Eric~P Xing.
\newblock {Alpa: Automating inter-and {Intra-Operator} parallelism for distributed deep learning.}
\newblock \emph{16th USENIX Symposium on Operating Systems Design and Implementation (OSDI 22)}, 2022.

\bibitem[Zheng et~al.(2025{\natexlab{a}})Zheng, Bao, Hou, Zheng, Fang, Huang, Li, Duanmu, Chen, Xu, Guo, Zheng, Jiang, Di, Wang, Ye, Lin, Chang, Lu, Liang, Zhai, and Liu]{25-triton-dist}
Size Zheng, Wenlei Bao, Qi~Hou, Xuegui Zheng, Jin Fang, Chenhui Huang, Tianqi Li, Haojie Duanmu, Renze Chen, Ruifan Xu, Yifan Guo, Ningxin Zheng, Ziheng Jiang, Xinyi Di, Dongyang Wang, Jianxi Ye, Haibin Lin, Li-Wen Chang, Liqiang Lu, Yun Liang, Jidong Zhai, and Xin Liu.
\newblock {Triton-distributed: Programming Overlapping Kernels on Distributed AI Systems with the Triton Compiler}.
\newblock \emph{arXiv preprint arXiv:2504.19442}, June 2025{\natexlab{a}}.

\bibitem[Zheng et~al.(2025{\natexlab{b}})Zheng, Fang, Zheng, Hou, Bao, Zheng, Jiang, Wang, Ye, Lin, Chang, and Liu]{25-tilelink}
Size Zheng, Jin Fang, Xuegui Zheng, Qi~Hou, Wenlei Bao, Ningxin Zheng, Ziheng Jiang, Dongyang Wang, Jianxi Ye, Haibin Lin, Li-Wen Chang, and Xin Liu.
\newblock {TileLink: Generating Efficient Compute-Communication Overlapping Kernels using Tile-Centric Primitives}.
\newblock \emph{arXiv preprint arXiv:2503.20313}, March 2025{\natexlab{b}}.

\bibitem[Zhu et~al.(2025)Zhu, Gao, Zhao, Zhao, Zuo, Gu, Xie, Tang, Xu, Ye, Kamahori, Lin, Wang, Wang, Krishnamurthy, and Kasikci]{25-nanoflow}
Kan Zhu, Yufei Gao, Yilong Zhao, Liangyu Zhao, Gefei Zuo, Yile Gu, Dedong Xie, Tian Tang, Qinyu Xu, Zihao Ye, Keisuke Kamahori, Chien-Yu Lin, Ziren Wang, Stephanie Wang, Arvind Krishnamurthy, and Baris Kasikci.
\newblock {NanoFlow: Towards Optimal Large Language Model Serving Throughput}.
\newblock \emph{arXiv preprint arXiv:2408.12757}, May 2025.

\end{thebibliography}

\appendix
\clearpage
\section*{Appendix}

We present {\name} performance on Blackwell GPUs (Appendix~\ref{appendix-blackwell-perf}), additional collective performance results (Appendix~\ref{appendix-additional-collective-perf}), {\name} API specification (Appendix~\ref{appendix-api}), program template and example kernels (Appendix~\ref{appendix-template-kernels}), multi-GPU setup process (Appendix~\ref{appendix-multi-gpu-setup-process}), and in-network acceleration setup process (Appendix~\ref{appendix-in-network-accel-setup-process}).

\section{Blackwell GPU Performance}
\label{appendix-blackwell-perf}

In this section, we demonstrate that {\shortname} generalizes across different hardware architectures by presenting representative kernel performance on Blackwell GPUs and comparing against available baselines that also support this architecture.

All experiments were conducted using \(8\times\)Nvidia B200 GPUs, interconnected via 5th-generation NVLink and NVSwitch (900 GB/s unidirectional bandwidth), using CUDA~12.8 and PyTorch~2.8.0. All matrix multiplications use BF16 as the element type and FP32 as the tensor core accumulator type. For brevity, we denote the GEMM shape as $M \times N \times K$, where the first operand has dimensions $M \times K$ and the second has dimensions $K \times N$. We report the observed average compute throughput.

\begin{figure}[H]
    \centering
    \includegraphics[
        width=0.6\textwidth
    ]{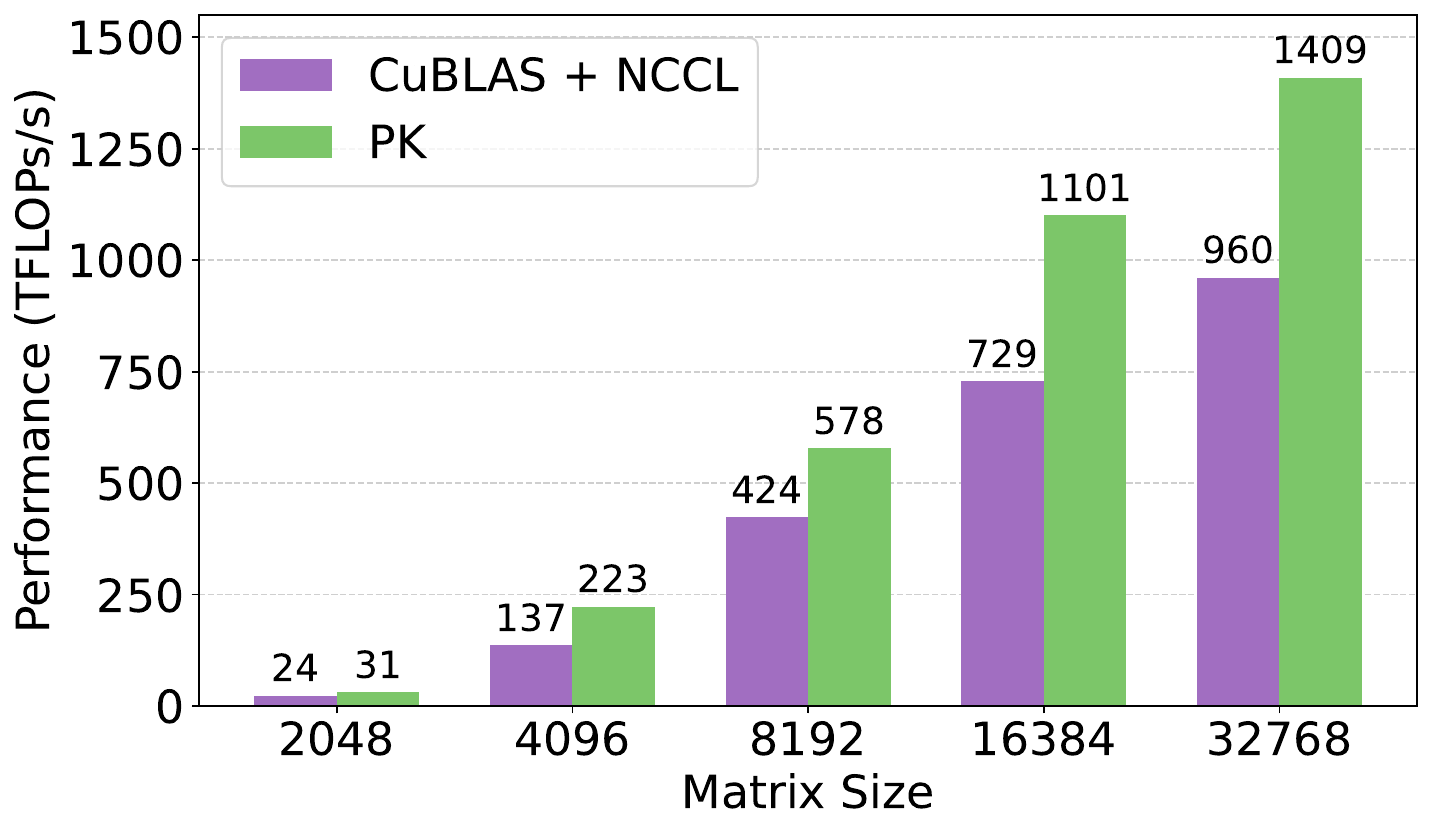}
    \caption{GEMM + RS performance. Local GEMM size is $N \times N \times N/8$, with $N$ given in the X-axis.}
    \label{fig:gemm-rs-b200}
\end{figure}

\begin{figure}[H]
    \centering
    \includegraphics[
        width=0.7\textwidth
    ]{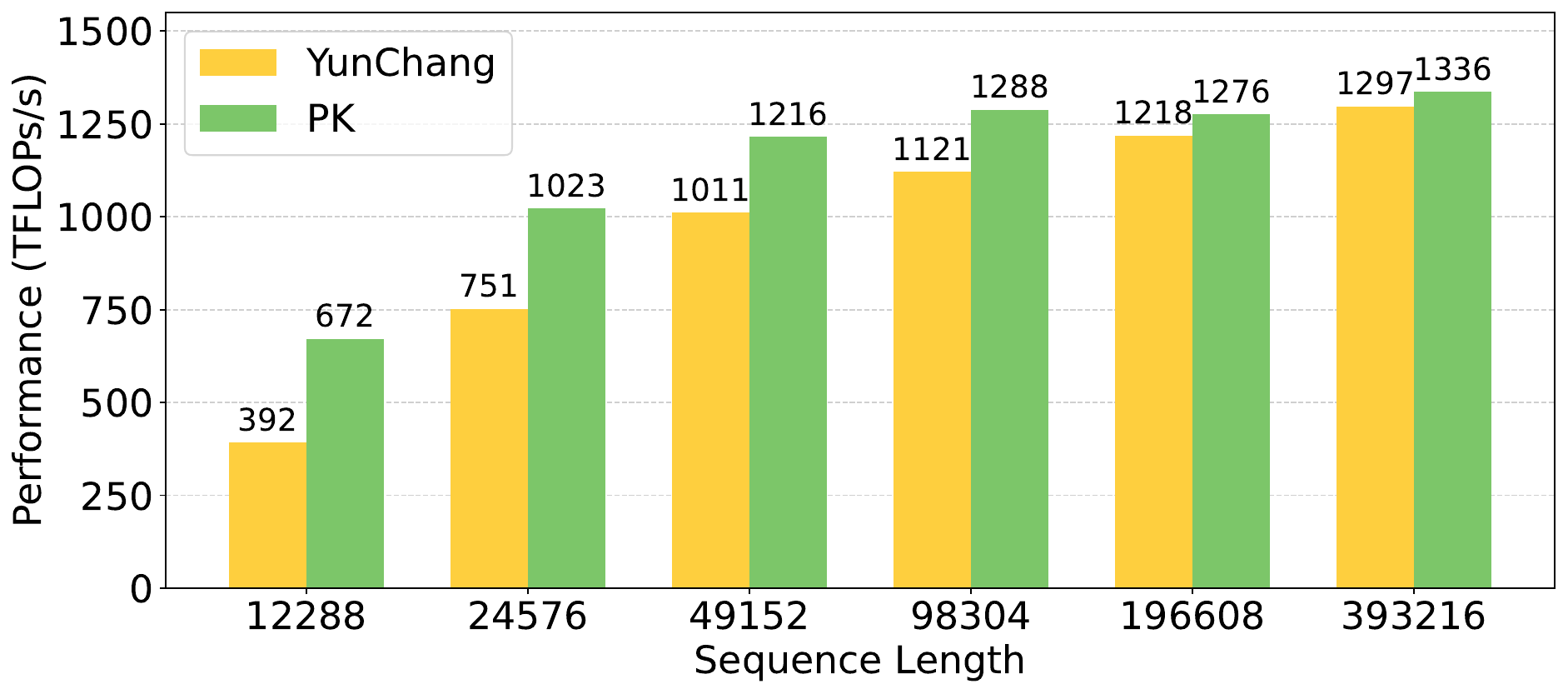}
    \caption{DeepSpeed-Ulysses attention layer performance across sequence lengths ($B=16$, $H=128$, $D=128$).}
    \label{fig:ulysses-attention-b200}
\end{figure}

\section{Additional Collective Performance}
\label{appendix-additional-collective-perf}

In this section, we report additional results on pure collective kernel performance and compare them against NCCL. We particularly examine how performance can improve significantly when the communication pattern is \textit{fine-grained}: for example, when performing all-gather or reduce-scatter along the tensor dimension (the last dimension) instead of the batch dimension (the first dimension), or when performing all-to-all operations across head and sequence dimensions. In such cases, the memory layout becomes discontiguous, which makes NCCL inefficient, as it supports collectives only on contiguous partitions and thus requires extra reshaping and copying. In contrast, {\shortname} can execute these collectives directly on the original layout. The results below illustrate this advantage.

\begin{figure}[H]
    \centering
    \includegraphics[
        width=0.6\textwidth
    ]{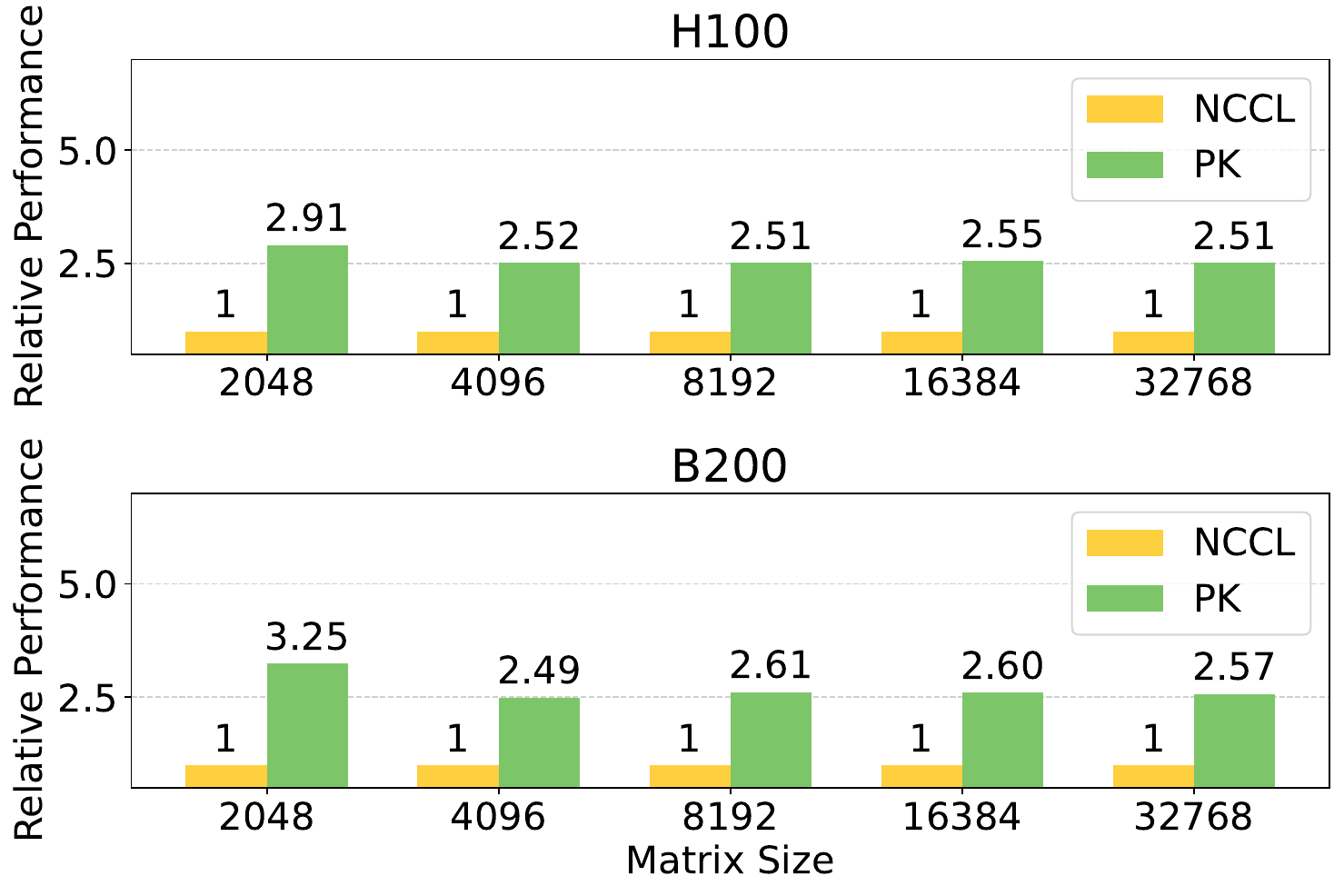}
    \caption{Tensor dimension all-gather performance comparison (BF16). The gathered matrix size is $N \times N$, with $N$ given in the X-axis.}
    \label{fig:all-gather}
\end{figure}

\begin{figure}[H]
    \centering
    \includegraphics[
        width=0.6\textwidth
    ]{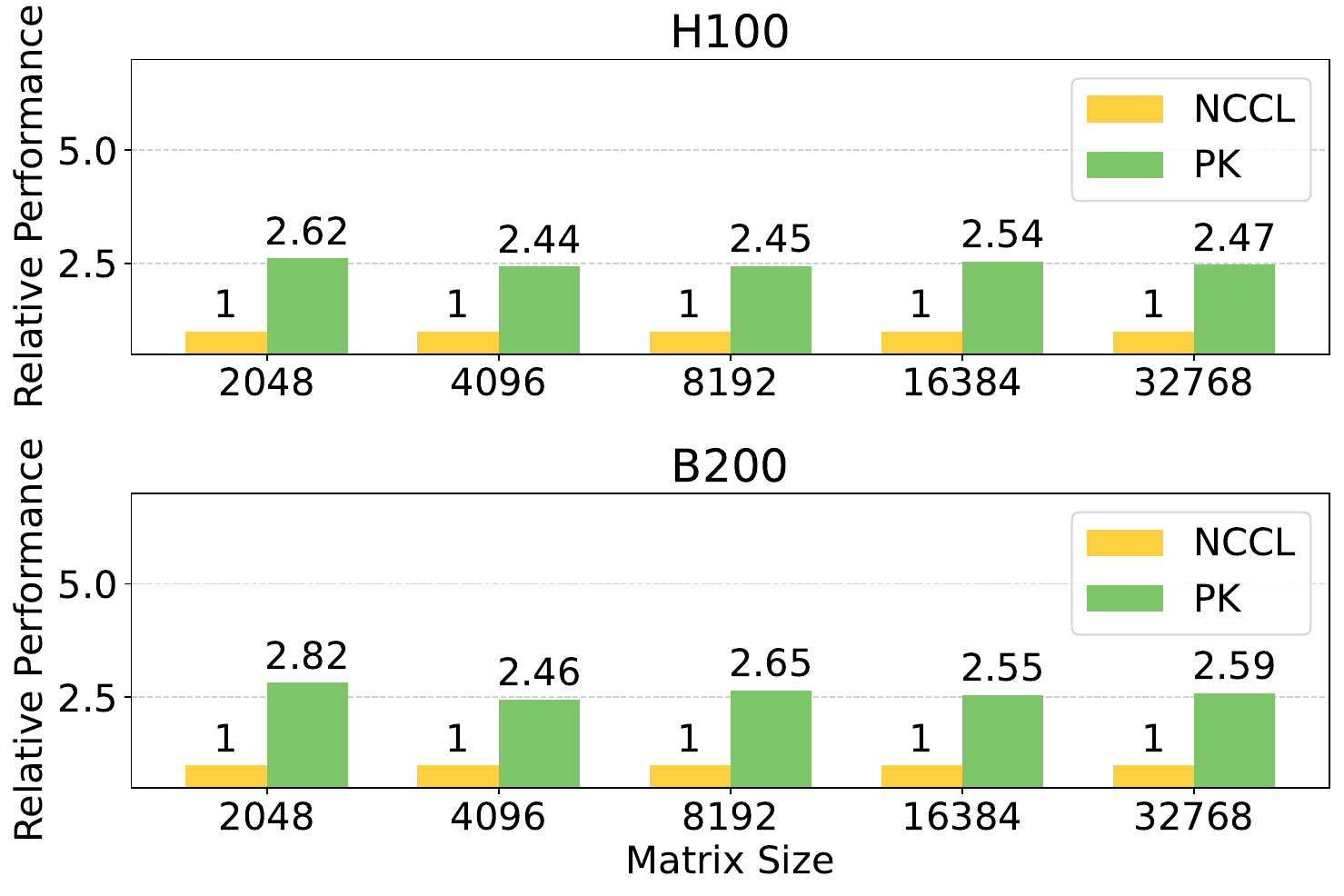}
    \caption{Tensor dimension reduce-scatter performance comparison (BF16). The scattered matrix size is $N \times N/8$, with $N$ given in the X-axis.}
    \label{fig:reduce-scatter}
\end{figure}

\begin{figure}[H]
    \centering
    \includegraphics[
        width=0.6\textwidth
    ]{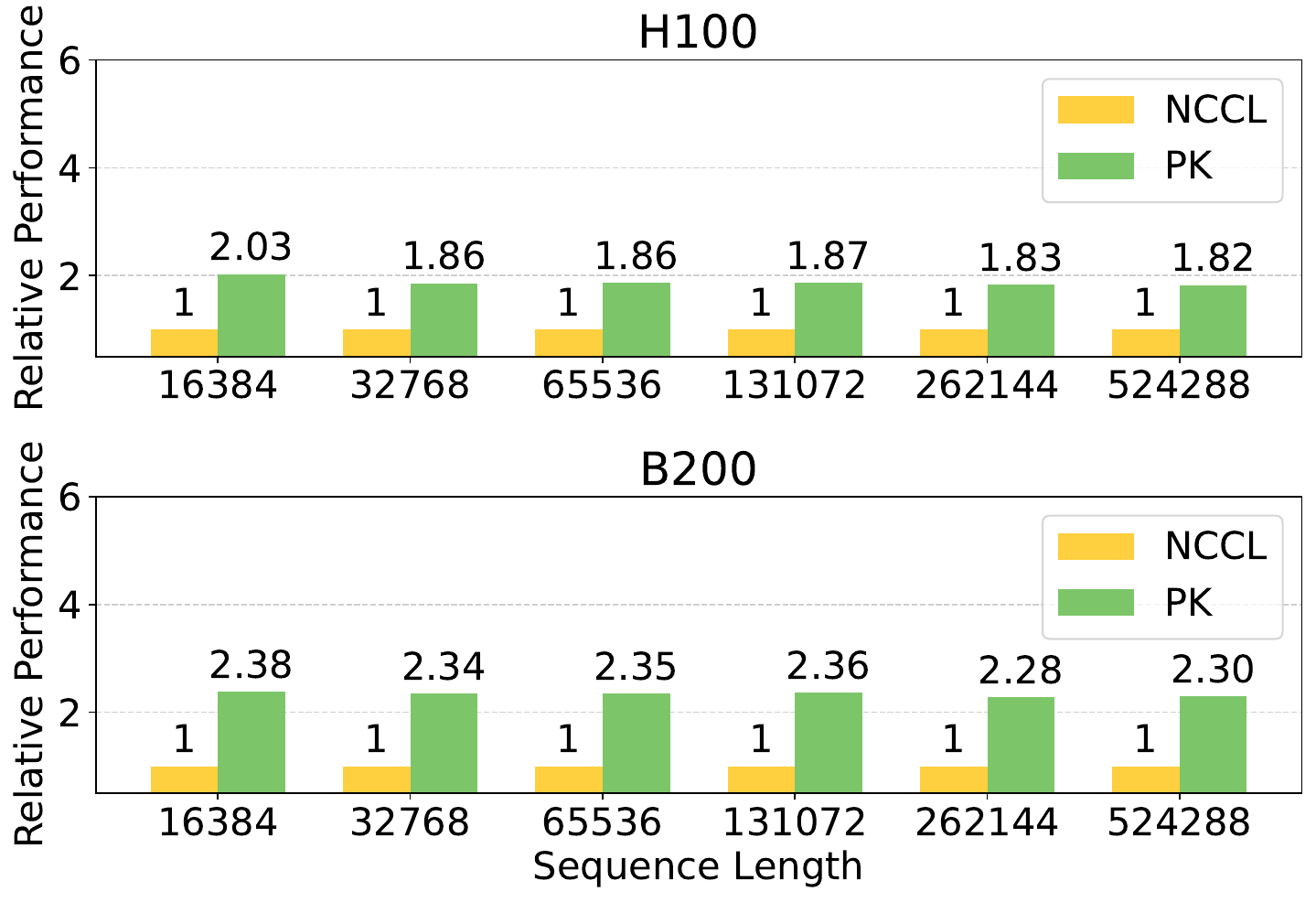}
    \caption{4-dimensional ($B$, $S$, $H$, $D$) all-to-all performance comparison (BF16), with $B=1$, $H=128$, $D=128$, and varying $S$ given in the X-axis. The $S$ dimension is gathered and the $H$ dimension is evenly scattered across 8 GPUs.}
    \label{fig:all-to-all}
\end{figure}

\section{{\name} API Specification}
\label{appendix-api}

We provide the full specification of {\shortname} primitives, including each function’s name, signature, parameters, and description.

\vspace{0.5em} \begin{verbatim}
template <int axis, cache_policy policy, kittens::ducks::st::all ST, 
          kittens::ducks::pgl::all PGL, kittens::ducks::coord::tile COORD>
__device__ void store_async(const PGL &dst, const ST &src, const COORD &idx)
\end{verbatim}
\vspace{0.5em} \textit{Template Parameters:} \vspace{0.3em}
\begin{itemize}[itemsep=0pt,topsep=0pt,leftmargin=*]
    \item \texttt{axis}: Tensor axis for the operation (0-3).
    \item \texttt{policy}: Cache policy (NORMAL or cache hint).
    \item \texttt{ST}: Shared tile type.
    \item \texttt{PGL}: Parallel global layout type.
    \item \texttt{COORD}: Coordinate type for indexing.
\end{itemize}
\vspace{1em} \textit{Parameters:} \vspace{0.3em}
\begin{itemize}[itemsep=0pt,topsep=0pt,leftmargin=*]
    \item \texttt{dst}: Destination parallel global layout.
    \item \texttt{src}: Source shared memory tile.
    \item \texttt{idx}: Coordinate specifying the destination position.
\end{itemize}
\vspace{1em} \textit{Description:}
Asynchronously stores a shared memory tile to multicast memory using the Tensor Memory Accelerator (TMA). Launched by a single thread.

\vspace{0.5em} \begin{verbatim}
template <int axis, cache_policy policy, kittens::ducks::st::all ST, 
          kittens::ducks::pgl::all PGL, kittens::ducks::coord::tile COORD>
__device__ void store_add_async(const PGL &dst, const ST &src, const COORD &idx)
\end{verbatim}
\vspace{0.5em} \textit{Template Parameters:} \vspace{0.3em}
\begin{itemize}[itemsep=0pt,topsep=0pt,leftmargin=*]
    \item \texttt{axis}: Tensor axis for the operation (0-3).
    \item \texttt{policy}: Cache policy (NORMAL or cache hint).
    \item \texttt{ST}: Shared tile type.
    \item \texttt{PGL}: Parallel global layout type.
    \item \texttt{COORD}: Coordinate type for indexing.
\end{itemize}
\vspace{1em} \textit{Parameters:} \vspace{0.3em}
\begin{itemize}[itemsep=0pt,topsep=0pt,leftmargin=*]
    \item \texttt{dst}: Destination parallel global layout.
    \item \texttt{src}: Source shared memory tile.
    \item \texttt{idx}: Coordinate specifying the destination position.
\end{itemize}
\vspace{1em} \textit{Description:}
Asynchronously performs an atomic add reduction from a shared memory tile to multicast memory via TMA. The operation atomically adds the source tile values to the existing values at the destination. Launched by a single thread.

\vspace{0.5em} \begin{verbatim}
template <int TILE_ROWS, int TILE_COLS, kittens::reduce_op OP, 
          kittens::ducks::pgl::all PGL, kittens::ducks::gl::all GL>
__device__ void reduce(GL &dst, const coord &dst_idx, PGL &src, const coord &src_idx)
\end{verbatim}
\vspace{0.5em} \textit{Template Parameters:} \vspace{0.3em}
\begin{itemize}[itemsep=0pt,topsep=0pt,leftmargin=*]
    \item \texttt{TILE\_ROWS}: Number of rows in the tile.
    \item \texttt{TILE\_COLS}: Number of columns in the tile.
    \item \texttt{OP}: Reduction operation to apply (sum, max, or min).
    \item \texttt{PGL}: Parallel global layout type.
    \item \texttt{GL}: Global layout type.
\end{itemize}
\vspace{1em} \textit{Parameters:} \vspace{0.3em}
\begin{itemize}[itemsep=0pt,topsep=0pt,leftmargin=*]
    \item \texttt{dst}: Reference to the destination global layout.
    \item \texttt{dst\_idx}: Coordinate specifying the destination tile's position.
    \item \texttt{src}: Reference to the source parallel global layout.
    \item \texttt{src\_idx}: Coordinate specifying the source tile's position.
\end{itemize}
\vspace{1em} \textit{Description:}
Performs a reduction from multicast memory to device-local global memory. The function loads data from the source parallel global layout using in-network reduction operations and stores the result to the destination global layout. Collectively launched by one or more warps. Each warp processes multiple rows of the tile, performing the specified reduction operation during the multicast load and then writing the reduced values to the destination global memory.

\vspace{0.5em} \begin{verbatim}
template <int TILE_ROWS, int TILE_COLS, kittens::reduce_op OP, 
          kittens::ducks::pgl::all PGL>
__device__ void all_reduce(PGL &dst_and_src, const coord &idx)
\end{verbatim}
\vspace{0.5em} \textit{Template Parameters:} \vspace{0.3em}
\begin{itemize}[itemsep=0pt,topsep=0pt,leftmargin=*]
    \item \texttt{TILE\_ROWS}: Number of rows in the tile.
    \item \texttt{TILE\_COLS}: Number of columns in the tile.
    \item \texttt{OP}: Reduction operation to apply (sum, max, or min).
    \item \texttt{PGL}: Parallel global layout type.
\end{itemize}
\vspace{1em} \textit{Parameters:} \vspace{0.3em}
\begin{itemize}[itemsep=0pt,topsep=0pt,leftmargin=*]
    \item \texttt{dst\_and\_src}: Reference to the parallel global layout object.
    \item \texttt{idx}: Coordinate specifying the tile's position with batch (\texttt{b}), depth (\texttt{d}), row (\texttt{r}), and column (\texttt{c}) indices.
\end{itemize}
\vspace{1em} \textit{Description:}
Performs an all-reduce collective operation on a tile of data on multicast memory. The function reduces data across all participating GPUs for the specified tile. Collectively launched by one or more warps. Each warp processes multiple rows, loading data from multicast memory with the specified reduction operation, then writing the result back to the same multicast location. The operation leverages in-network acceleration hardware to efficiently perform the reduction without explicit peer-to-peer copies.

\vspace{0.5em} \begin{verbatim}
__device__ void signal(const barrier_t &barrier, const coord &idx, 
                       const int dst_dev_idx, const int val)
\end{verbatim}
\vspace{0.5em} \textit{Parameters:} \vspace{0.3em}
\begin{itemize}[itemsep=0pt,topsep=0pt,leftmargin=*]
    \item \texttt{barrier}: Reference to the barrier object (parallel global layout of integers).
    \item \texttt{idx}: Element-wise coordinate specifying the barrier location.
    \item \texttt{dst\_dev\_idx}: Target device index to signal.
    \item \texttt{val}: Value to add to the barrier counter.
\end{itemize}
\vspace{1em} \textit{Description:}
Signals a specific device's barrier by atomically adding a value to its counter. This primitive is used to coordinate synchronization between thread blocks and GPUs.

\vspace{0.5em} \begin{verbatim}
__device__ void signal_all(const barrier_t &barrier, const coord &idx, const int val)
\end{verbatim}
\vspace{0.5em} \textit{Parameters:} \vspace{0.3em}
\begin{itemize}[itemsep=0pt,topsep=0pt,leftmargin=*]
    \item \texttt{barrier}: Reference to the barrier object.
    \item \texttt{idx}: Element-wise coordinate specifying the barrier location.
    \item \texttt{val}: Value to add to all devices' barrier counters.
\end{itemize}
\vspace{1em} \textit{Description:}
Signals all devices simultaneously by performing a multicast atomic add operation. Uses in-network multicast hardware to efficiently update barrier counters across all participating devices with a single operation.

\vspace{0.5em} \begin{verbatim}
__device__ void wait(const barrier_t &barrier, const coord &idx, 
                     const int dev_idx, const int expected)
\end{verbatim}
\vspace{0.5em} \textit{Parameters:} \vspace{0.3em}
\begin{itemize}[itemsep=0pt,topsep=0pt,leftmargin=*]
    \item \texttt{barrier}: Reference to the barrier object.
    \item \texttt{idx}: Element-wise coordinate specifying the barrier location.
    \item \texttt{dev\_idx}: Device index to wait on.
    \item \texttt{expected}: Expected barrier value to wait for.
\end{itemize}
\vspace{1em} \textit{Description:}
Waits until a device's barrier counter reaches the expected value. Continuously polls the barrier location using relaxed memory ordering loads until the expected value is observed. This provides a spinning wait mechanism for inter-SM and inter-GPU synchronization.

\vspace{0.5em} \begin{verbatim}
__device__ void barrier(const barrier_t &barrier, const coord &idx, const int dev_idx)
\end{verbatim}
\vspace{0.5em} \textit{Parameters:} \vspace{0.3em}
\begin{itemize}[itemsep=0pt,topsep=0pt,leftmargin=*]
    \item \texttt{barrier}: Reference to the barrier object.
    \item \texttt{idx}: Element-wise coordinate specifying the barrier location.
    \item \texttt{dev\_idx}: Current device index.
\end{itemize}
\vspace{1em} \textit{Description:}
Implements a complete barrier synchronization across all devices. This ensures all participating GPUs reach the same synchronization point before proceeding.

\section{{\name} Program Template and Example Kernels}
\label{appendix-template-kernels}

\paragraph{Load-Compute-Store-Communicate (LCSC) Template.}

The LCSC template provides a structured approach for implementing multi-GPU kernels with specialized worker components. The template enables flexible warp/SM specialization and overlapping strategies for compute, memory, and communication operations.

\vspace{1em} \noindent \textit{High-level Template Structure:} \vspace{0.3em}
\begin{verbatim}
struct lcsc_template {
    static void loader(globals, comp_sem, comp_smem, comp_regs);
    static void storer(globals, comp_sem, comp_smem, comp_regs);
    static void consumer(globals, comp_sem, comp_smem, comp_regs);
    static void communicator(globals, comm_sem, comm_smem, comm_regs);
};
\end{verbatim}

\vspace{1em} \noindent \textit{Required Components:} \vspace{0.3em}
\begin{itemize}[itemsep=0pt,topsep=0pt,leftmargin=*]
    \item \texttt{comp\_sem}: struct of semaphores for synchronization within compute SMs.
    \item \texttt{comm\_sem}: struct of semaphores for synchronization within communication SMs.
    \item \texttt{comp\_smem}: struct of shared memory layouts for compute SMs.
    \item \texttt{comm\_smem}: struct of shared memory layouts for communication SMs.
    \item \texttt{comp\_regs}: struct of register state for compute workers.
    \item \texttt{comm\_regs}: struct of register state for communication workers.
\end{itemize}

\vspace{1em} \noindent \textit{Workers:} \vspace{0.3em}
\begin{itemize}[itemsep=0pt,topsep=0pt,leftmargin=*]
    \item \texttt{loader}: Performs memory loads from local or peer HBM using TMA.
    \item \texttt{storer}: Performs memory stores to local or peer HBM.
    \item \texttt{consumer}: Performs tensor/CUDA core operations on loaded data.
    \item \texttt{communicator}: Performs dedicated inter-GPU communication. Executes on separate communication SMs.
\end{itemize}

\vspace{1em} \noindent \textit{Execution Model:}
The template automatically distributes SMs between computation and communication roles based on \texttt{num\_comm\_sms}, passed in to the host entry function. Compute SMs execute loader, storer, and consumer functions with producer-consumer synchronization through semaphores. Communication SMs execute the communicator function independently. The framework handles warpgroup specialization, register allocation, and task distribution across workers. Programmers can utilize this template by defining the above struct, and passing it to the launch interface:

\begin{verbatim}
lcsc::launch_kernel<config, globals, lcsc_template>(G, stream);
\end{verbatim}

\noindent Where the parameters are: \vspace{0.3em}

\begin{itemize}[itemsep=0pt,topsep=0pt,leftmargin=*]
    \item \texttt{config}: Compile-time configuration struct defining SM and thread counts.
    \item \texttt{globals}: Runtime globals struct containing device memory pointers and parameters.
    \item \texttt{lcsc\_template}: User-defined LCSC template implementation.
    \item \texttt{G}: Instance of globals struct.
    \item \texttt{stream}: CUDA stream for kernel execution.
\end{itemize}

\vspace{0.5em} We present a fused GEMM + all-reduce (AR) kernel implemented using the LCSC template in Figure~\ref{fig:lcsc-gemm-ar}. We highlight that the kernel contains \textit{both} a fully optimized GEMM and fused all-reduce logic, with the communication-relevant code comprising only about 10 lines of device code. We also open-source all remaining kernels evaluated in this paper through our GitHub repository.

\begin{figure}[H]
\begin{lstlisting}[basicstyle=\scriptsize\ttfamily]
__device__ inline void loader(const globals &G, comp_sem &sem, comp_smem &smem, comp_regs &regs) { 
    int2 idx = interpret_task(regs.task_id);
    for (int red_idx = 0; red_idx < regs.num_iters; red_idx++) {
        wait(sem.inputs_finished[regs.stage], get_phasebit<1>(regs.phasebits, regs.stage));
        update_phasebit<1>(regs.phasebits, regs.stage);
        tma::expect_bytes(sem.inputs_arrived[regs.stage], sizeof(A_tile) * 2 + sizeof(B_tile));
        if (red_idx == PIPELINE_STAGES - 1) {
            wait(sem.outputs_finished, get_phasebit<1>(regs.phasebits, PIPELINE_STAGES));
            update_phasebit<1>(regs.phasebits, PIPELINE_STAGES);
        }
        for (int i = 0; i < 2; i++)
            tma::load_async(smem.inputs[regs.stage].A[i], G.A, {idx.x * 2 + i, red_idx}, sem.inputs_arrived[regs.stage]);
        tma::load_async(smem.inputs[regs.stage].B, G.B, {red_idx, idx.y}, sem.inputs_arrived[regs.stage]);
        regs.stage = (regs.stage + 1) % PIPELINE_STAGES;
    }
}

__device__ inline void storer(const globals &G, comp_sem &sem, comp_smem &smem, comp_regs &regs) { 
    int2 idx = interpret_task(regs.task_id);
    wait(sem.outputs_arrived, get_phasebit<0>(regs.phasebits, 0));
    update_phasebit<0>(regs.phasebits, 0);
    for (int i = 0; i < 2; i++)
        tma::store_async(G.C[G.dev_idx], regs.C[i], {idx.x * 2 + i, idx.y});
    tma::store_async_read_wait();
    arrive(sem.outputs_finished);
    int signal_dev_idx = regs.task_id % NUM_DEVICES;
    device<NUM_DEVICES>::signal(G.barrier, {idx.x, idx.y}, signal_dev_idx, 1);
}

__device__ inline void consumer(const globals &G, comp_sem &sem, comp_smem &smem, comp_regs &regs) {
    rt_fl<ROW_BLOCK / 8, COL_BLOCK> C_accum;
    warp::zero(C_accum);
    for (int red_idx = 0; red_idx < regs.num_iters; red_idx++) {
        wait(sem.inputs_arrived[regs.stage], get_phasebit<0>(regs.phasebits, regs.stage));
        update_phasebit<0>(regs.phasebits, regs.stage);
        warpgroup::mma_AB(C_accum, smem.inputs[regs.stage].A[regs.warpgroup_id], smem.inputs[regs.stage].B);
        warpgroup::mma_async_wait();
        warp::arrive(sem.inputs_finished[regs.stage]);
        regs.stage = (regs.stage + 1) % PIPELINE_STAGES;
    }
    group<8>::sync(3);
    warpgroup::store(regs.C[regs.warpgroup_id], C_accum);
    warpgroup::sync(regs.warpgroup_id + 1);
    warpgroup::arrive(sem.outputs_arrived);
}

__device__ inline void communicator(const globals &G, comm_sem &sem, comm_smem &smem, comm_regs &regs) {
    int2 idx = interpret_task(regs.task_id);
    if (threadIdx.x == 0)
        device<NUM_DEVICES>::wait(G.barrier, {idx.x, idx.y}, G.dev_idx, NUM_DEVICES);
    __syncthreads();
    group<NUM_WARPS>::all_reduce<ROW_BLOCK, COL_BLOCK, reduce_op::ADD>(G.C, {idx.x, idx.y});
}
\end{lstlisting}
\caption{Fused GEMM + AR kernel implemented with the LCSC template}
\label{fig:lcsc-gemm-ar}
\end{figure}

\section{Multi-GPU Memory Setup Process}
\label{appendix-multi-gpu-setup-process}

We describe the low-level multi-GPU memory setup process, a major complexity in multi-GPU programming, which {\shortname} abstracts away from programmers.

The basic requirement of multi-GPU programming is that kernels must be able to access memory (HBM) on peer devices. To enable this, we need to create a new mapping in the current device’s virtual address space that points to the peer device’s physical memory. After such, the kernel can simply dereference the address, and the NVLink and NVSwitch fabric handle the underlying transfer.

There are three ways to create such mappings: (1) CUDA Unified Virtual Addressing, (2) CUDA Inter-Process Communication, and (3) manual Virtual Memory Management.

\subsection{CUDA Unified Virtual Addressing (UVA)}

UVA provides a single unified virtual address space across GPUs, but with the limitation that it applies only within a single process. That is, if we avoid using multiple processes altogether, there exists no heterogeneous virtual address spaces.

\begin{figure}[t]
    \centering
    \includegraphics[
        width=0.6\textwidth
    ]{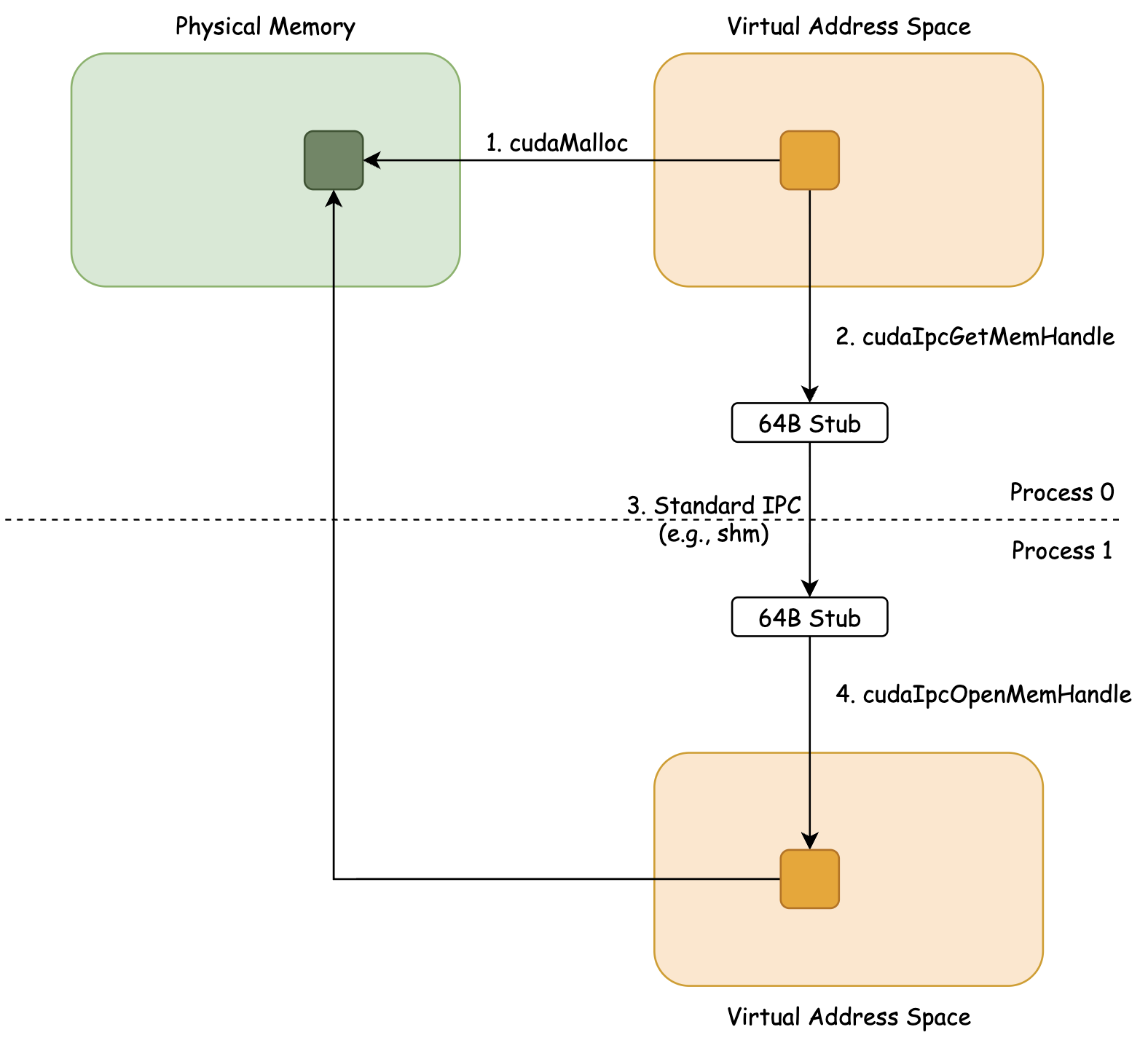}
    \caption{CUDA IPC flow.}
    \label{fig:cuda-ipc-flow}
\end{figure}

However, we note that modern production distributed training and inference are built around a multi-processing model. Distributed runners like \texttt{torchrun} assume 1 GPU device per rank (process), and working around this is quite complicated. Thus, multi-processing is the preferred model of launching multi-GPU workloads, which brings us to the next two methods.

\subsection{CUDA Inter-Process Communication (IPC)}

Calling \texttt{cudaIpcGetMemHandle} on the address in the current virtual address space returns a 64-byte stub that can be shared across processes through standard IPC mechanisms like shared memory or Unix domain sockets. The receiving process then can call \texttt{cudaIpcOpenMemHandle}, which maps the given stub into its own address space. Figure~\ref{fig:cuda-ipc-flow} visualizes this flow.

While this method is straightforward and works on pre-allocated device memory (e.g., existing PyTorch tensors), its drawback is that it cannot use the NVSwitch accelerator for faster reduction and broadcast operations.

\begin{figure}[t]
    \centering
    \includegraphics[
        width=0.6\textwidth
    ]{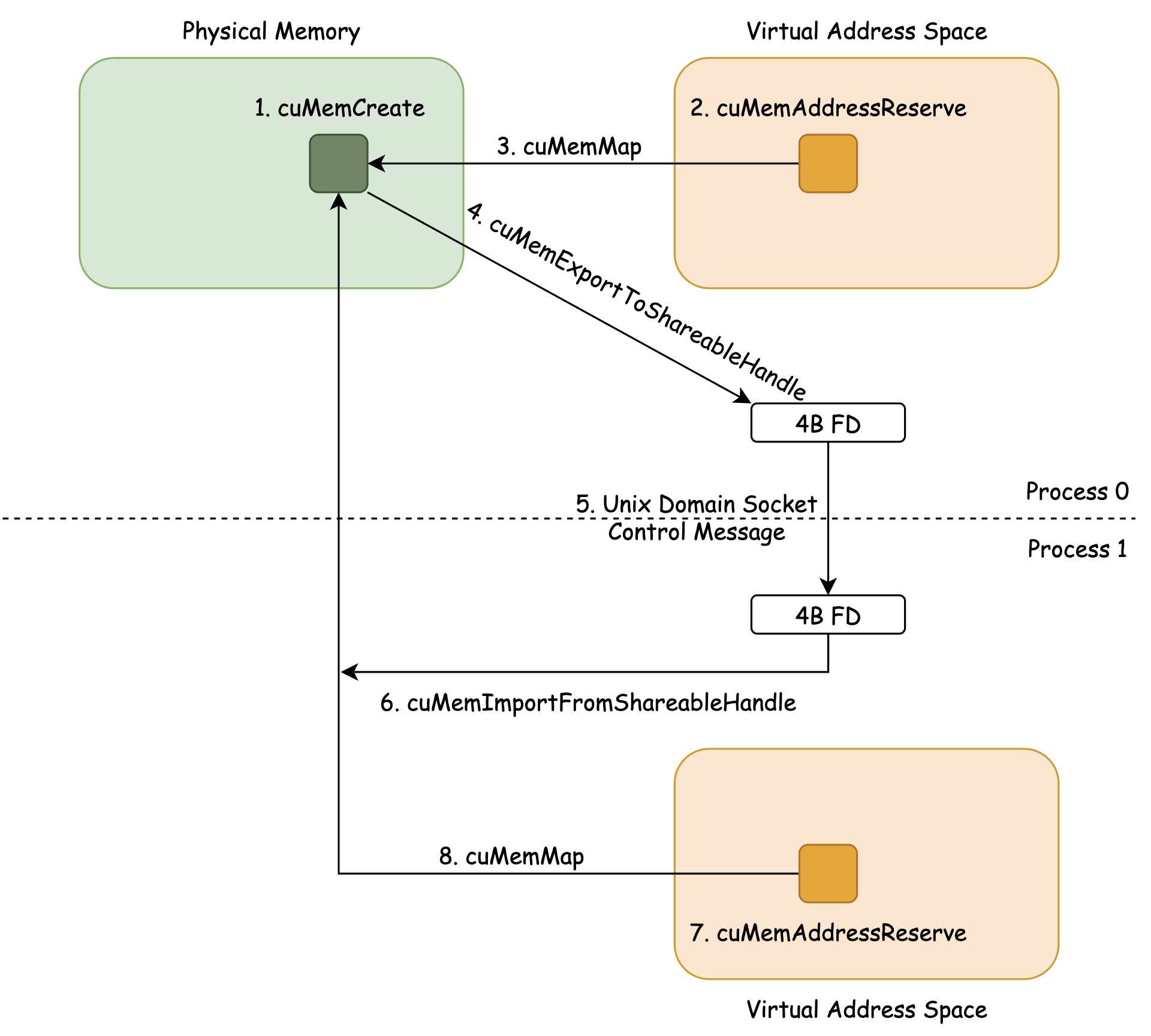}
    \caption{CUDA VMM flow.}
    \label{fig:vmm-flow}
\end{figure}

\begin{figure}[t]
    \centering
    \includegraphics[
        width=0.6\textwidth
    ]{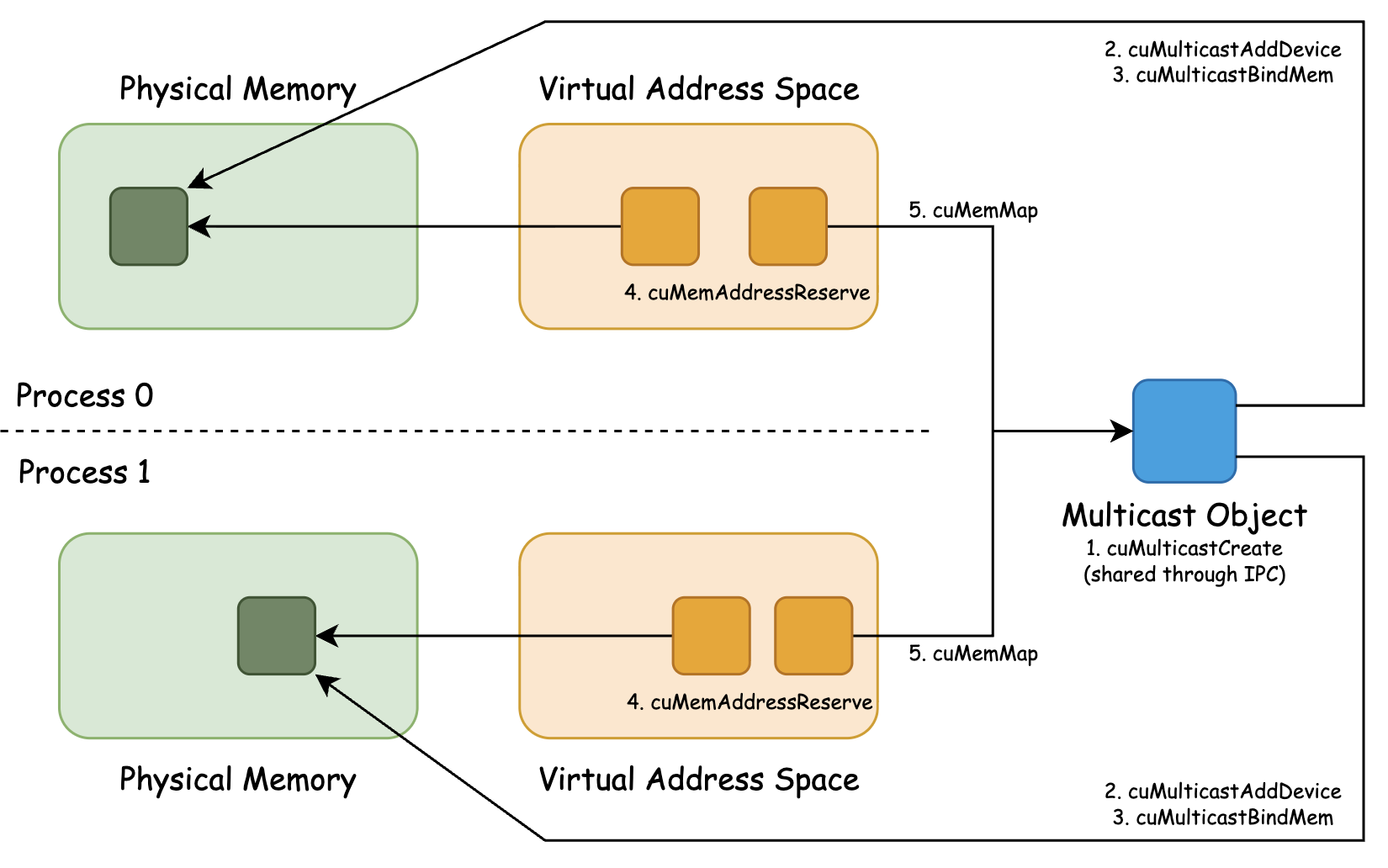}
    \caption{CUDA multicast object creation process.}
    \label{fig:multicast-creation}
\end{figure}

\subsection{Manual Virtual Memory Management (VMM)}

For VMM, we start by manually allocating the GPU physical memory with \texttt{cuMemCreate}. This allows setting the \texttt{CU\_MEM\_HANDLE\_TYPE\_POSIX\_FILE\_DESCRIPTOR} property on this physical memory, which then lets us export the physical memory reference as a Linux file descriptor by calling \texttt{cuMemExportToShareableHandle}.

Because file descriptors are tied to a specific process in Linux, they cannot be shared directly. The standard way to transfer a file descriptor in Linux is to send it as a control message over a Unix domain socket. Once we send the file descriptor over to the destination process, it can then import the physical memory reference using \texttt{cuMemImportFromShareableHandle} and map it into its own virtual address space using the VMM API. The overall flow is illustrated in Figure~\ref{fig:vmm-flow}.

A downside of this approach is that the given memory must be allocated with VMM and is subject to size granularity requirements, typically at 2MB for H100s and B200s. As a result, a PyTorch-allocated tensor, which is usually allocated by the standard cudaMalloc without size alignment, cannot be shared directly across processes. Instead, we need a custom tensor class that manages device memory allocation and deallocation with custom VMM logic. The main advantage, however, is that this method enables the use of NVSwitch in-network accelerators.

\section{In-network Acceleration Setup Process}
\label{appendix-in-network-accel-setup-process}

In order to utilize NVSwitch acceleration, we first allocate local memory on each participating device with VMM. Then we create a \textit{multicast object}, which is an abstraction over multiple physical locations in multiple devices. To do this, we create a 8-byte stub that represents the multicast object with \texttt{cuMulticastCreate}, register all devices as participants, and map each device’s physical memory region to it.

A multicast object behaves just like VMM-allocated physical memory: we can share it with other processes and map a virtual address to it using the same mechanism described in the VMM setup process. That is, we export the multicast object as a POSIX file descriptor, open them on each device, and map them into each process’s virtual address space. The overall setup process and the exact names of the CUDA functions called are shown in Figure~\ref{fig:multicast-creation}.

\begin{figure}[H]
    \centering
    \includegraphics[
        width=0.6\textwidth
    ]{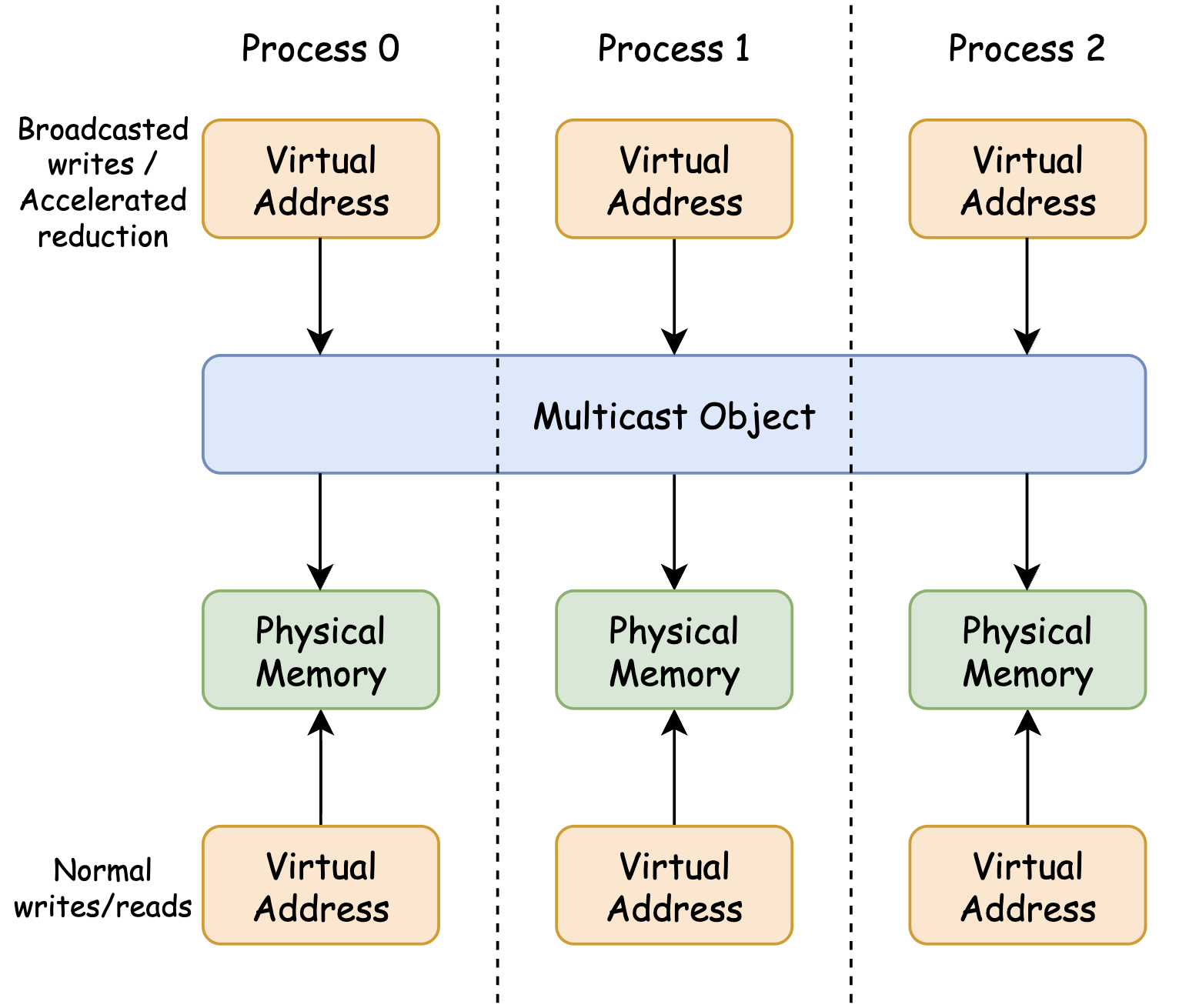}
    \caption{CUDA multicast object hierarchy.}
    \label{fig:multicast-hierarchy}
\end{figure}

After completing the above, each process has two addresses: one mapping to the current device’s physical memory (local address) and another mapping to the multicast object (multicast address). Writing to and reading from the local address is a standard global memory access. Writing to the multicast address triggers a broadcast across all participating devices, multicasted in the NVSwitch fabric. Reading from the multicast address causes undefined behavior. Finally, in-fabric reduction operations can be invoked on the multicast address using the PTX instructions \texttt{multimem.red} and \texttt{multimem.ld\_reduce}. This is illustrated in Figure~\ref{fig:multicast-hierarchy}.

\end{document}